\documentclass[prb,aps,twocolumn,, nofootinbib]{revtex4-2} 

\usepackage{amsmath}
\usepackage{amssymb}
\usepackage{mathrsfs}
\usepackage[bbgreekl]{mathbbol}
\usepackage{mathtools}
\usepackage{physics}
\usepackage[american]{babel}

\usepackage{multirow}
\usepackage{hhline}

\usepackage[x11names]{xcolor}
\usepackage[colorlinks=true, citecolor=blue!90!black, linkcolor=blue!90!black, linktocpage=true, urlcolor=red!70!black]{hyperref}

\usepackage{tikz}
\usetikzlibrary{decorations.pathreplacing}

\definecolor{MajBlue}{HTML}{377eb8}
\definecolor{MajRed}{HTML}{e41a1c}

\newcommand{\secref}[1]{Section~\ref{#1}}
\renewcommand{\eqref}[1]{Eq.~(\ref{#1})}
\newcommand{\appref}[1]{Appendix~\ref{#1}}
\newcommand{\figref}[1]{Fig.~\ref{#1}}
\newcommand{\tabref}[1]{Table~\ref{#1}}

\makeatletter
\newcommand{\HideAppendixSubsectionsFromToc}{%
  \let\l@subsection\@gobbletwo
  \let\l@subsubsection\@gobbletwo
}
\makeatother

\makeatletter
\newenvironment{widetextnorules}{%
  \par\ignorespaces
  \onecolumngrid
  \vskip10\p@
  \vskip6\p@
  \prep@math@patch
}{%
  \par
  \vskip6\p@
  \vskip8.5\p@
  \twocolumngrid\global\@ignoretrue
  \@endpetrue
}
\makeatother

\begin{document}

\title{Infinite-Order Lattice Chiral Anomalies and CPT
}

\hfill MIT-CTP/6050

\author{
Elijah Lew-Smith,\hspace{-0.5ex}\textsuperscript{$a_j$}
Salvatore D. Pace,\hspace{-0.5ex}\textsuperscript{$b_j$}
Shu-Heng Shao\textsuperscript{$a_j$}
}

\affiliation{\textsuperscript{$a_j$} Center for Theoretical Physics --- a Leinweber Institute, Massachusetts Institute of Technology\\
\textsuperscript{$b_j$} Department of Physics, Massachusetts Institute of Technology}

\begin{abstract}

A key property of a global symmetry's anomaly is its order: the smallest integer $n$ for which the diagonal symmetry of the $n$-copy system is anomaly-free. 
While many familiar lattice anomalies have finite order, perturbative anomalies in the continuum---those captured by Feynman diagrams---have infinite order.
In this paper, we show that the Onsager symmetry, a lattice realization of the chiral symmetry of a ${1+1}$d massless Dirac fermion, has an order-two anomaly. 
However, imposing lattice CPT symmetry enhances this anomaly from order two to infinite order, yielding a lattice chiral symmetry structure that more faithfully matches the continuum chiral anomaly. 
We also discuss the corresponding ${2+1}$d symmetry-protected topological phases for these infinite-order lattice anomalies.

\end{abstract}

\maketitle

\tableofcontents 

\section{Introduction}
\label{sec:introduction}

Anomalies of global symmetries play a central role in the study of quantum systems. 
They are part of the defining structural data of a quantum theory, and non-perturbatively constrain the possible IR dynamics. 
The theory of anomalies of ordinary global symmetries in quantum field theory (QFT) is by now mature and well developed. 
Recently, there has been growing interest in anomalies in quantum lattice systems and their relation to QFT anomalies in the IR limit (see~\cite{
CS221112543, 
S230805151, 
KS240102533, 
CPS240912220, 
PCS241218606, 
SS250309717,
BDD250315088,
PKC250504684, 
KX250504719, 
KS250707430,
KS250716966, 
TLE250721209,
SZJ250721267, 
SSZ250817115,
FKC250912304,
LYZ251006555,
FCH251023701,
CGT251202105,
SZ260101191,
PB260211266, 
L260213948,
ZC260400347} for a selection of recent work).

\begin{table*}[t]
\centering
\setlength{\tabcolsep}{8pt}
\renewcommand{\arraystretch}{1.5}

\begin{tabular}{|c|c|c|c|c|} 
\hline  
Model&
UV symmetry & IR symmetry & Anomaly order UV $\to$ IR & Reference \\ 
\hhline{|=|=|=|=|=|}
Two qubits
&$D_8$ & $\mathbb Z_2\times\mathbb Z_2$ & ${1\to 2}$ & \secref{anomQM} \\
\hline 
Compact bosons
&$\mathbb Z_4$ & $\mathbb Z_2$ & ${1\to 2}$ & \cite{TVV200806638} \\
\hline 
\multirow{2}{*}{\text{Heisenberg chain}}
&$\mathbb Z^\text{tran}$  & $\mathbb Z_2$ & ${1\to 2}$ & \cite{MT170707686,CS221112543}  \\
\cline{2-5}
&$\mathbb Z^\text{tran}\rtimes \mathbb Z_2^\text{CPT}$  & $\mathbb Z_2\times \mathbb Z_2^\text{CPT}$ & ${2\to 2}$ & \cite{SSZ250817115}  \\
\hline 
\multirow{2}{*}{\text{Kitaev chain}}&$\mathbb Z^\text{Maj}$ & $\mathbb Z_2$ & ${2\to 8}$ & \cite{SS230702534} \\
\cline{2-5}
&$\mathbb Z^\text{Maj}\rtimes \mathbb Z_2^\text{CPT}$  & $\mathbb Z_2 \times \mathbb Z_2^\text{CPT}$ & ${8\to 8}$ & \cite{SSZ250817115} \\
\hline
\multirow{3}{*}{\text{Staggered fermion}}& $\mathrm{U}(1)\times \mathbb{R}$  & 
\multirow{2}{*}{$\dfrac{\mathrm{U}(1)_\text{V} \times  \mathrm{U}(1)_\text{A}}{\mathbb Z_2}$}
& ${1\to \infty}$ & \secref{U(1)xR section} \\
\cline{2-2}
\cline{4-5}& Onsager  & 
& ${2\to \infty}$ & \secref{sec:lattice_anomalies} \\
\cline{2-5}& Onsager$\rtimes\mathbb Z_2^\text{CPT}$  & $\frac{\mathrm{U}(1)_\text{V} \times  \mathrm{U}(1)_\text{A}}{\mathbb Z_2} \times \mathbb Z_2^\text{CPT}$ & ${\infty\to \infty}$ & \secref{sec:lattice_crt} \\
\hline
\end{tabular}

\caption{\label{table} 
Examples of how the order of an anomaly changes from the UV to the IR. The symmetry groups $\mathbb Z^\text{tran}$ and $\mathbb Z^\text{Maj}$ denote the one-dimensional lattice translation groups of the Heisenberg spin chain and Kitaev's Majorana chain, respectively.
The IR symmetry column refers to the part of the IR symmetry that emanates from the UV symmetry, in the sense of~\cite{CS221112543}.
In examples with a lattice $\mathbb Z_2^\mathrm{CPT}$ CPT symmetry, enforcing CPT enhances the order of the UV anomaly to match the order of the IR anomaly.
In the compact bosons example, the flow is from two compact bosons to one, and is entirely in the continuum.
The flow in other examples is from a UV lattice system to an IR continuum field theory.
In the staggered fermion example, the IR $\mathrm{U}(1)_\text{V}$ and $ \mathrm{U}(1)_\text{A}$ symmetries are the vector and axial U(1) symmetries, respectively, of a Dirac fermion.
}
\renewcommand{\arraystretch}{1}
\end{table*}

\begin{figure}[t!]
\centering
\begin{tikzpicture}[
  line join=round,
  line cap=round,
  layeredge/.style={draw=SteelBlue4, line width=0.9pt},
  glabel/.style={font=\large, anchor=east},
  gaction/.style={->, draw=black!70, line width=0.8pt},
  scale=0.7
]
\definecolor{layerblue}{RGB}{164,215,243}
\def\layerwidth{4.9}
\def\layerdepth{1.25}
\def\layerskew{1.15}

\newcommand{\stacklayer}[2]{%
  \begin{scope}[shift={(#1,{#2-0.5})}]
    \coordinate (A) at (0,0);
    \coordinate (B) at (\layerwidth,0);
    \coordinate (C) at ({\layerwidth+\layerskew},\layerdepth);
    \coordinate (D) at (\layerskew,\layerdepth);
    \shade[
      layeredge,
      top color=layerblue!35!white,
      bottom color=layerblue
    ] (A) -- (B) -- (C) -- (D) -- cycle;
  \end{scope}
}
  \newcommand{\actionon}[2]{%
  \begin{scope}[shift={(#1,#2)}]
    \node[anchor=east] at (-0.4,0) () {$G$};
    \draw[samples=100,domain=45:315, shift={(0,0)}, line width=0.8pt] plot ({0.3*cos(\x)},{0.3*sin(\x)}); 
    \draw[<-, shift={(0,0)}, line width=0.8pt] ({0.3*cos(3+45)+0.1*sin(3+45)},{0.3*sin(3+45)-0.1*cos(3+45)})--++({-0.1*sin(3+45)},{0.1*cos(3+45)});
  \end{scope}
  }

  \actionon{-0.5}{-0.4}
  \actionon{-0.5}{0.6}
  \actionon{-0.5}{2.7}

\stacklayer{0}{0}
\stacklayer{0}{0.5}
\stacklayer{0}{2.5}

\node[font=\Large] at (3,1.72) {$\vdots$};
\draw[decorate, decoration={brace, amplitude=8pt}, line width=0.8pt]
  (6.3,3.5) -- (6.3,-0.4) node[midway, right=10pt] {$n$};
\end{tikzpicture}
\caption{The order of the anomaly for a global symmetry $G$ is defined as the smallest integer $n$ such that the diagonal $G$ symmetry in the $n$-copy system is anomaly-free.}
\label{fig:stacked-gc-layers}
\end{figure}
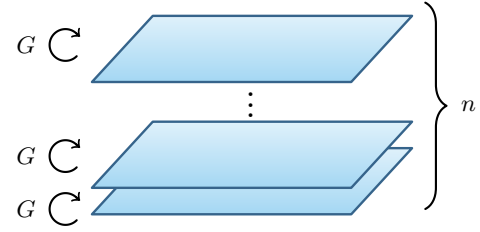

An important characteristic of an anomaly $\omega$ is its order, $\mathrm{ord}(\omega)$.
Consider a system with an anomalous unitary $G$ symmetry, where $G$ is a group and the symmetry operators are denoted by $\{U_g\}_{g\in G}$. 
We define the order of the anomaly by stacking identical copies of the system (see \figref{fig:stacked-gc-layers}).
If $\{U_g^{(i)}\}_{g\in G}$ denotes the symmetry operators of the $i$th copy, then the diagonal symmetry of $n$ copies is formed by ${\{\bigotimes_{i=1}^n U_g^{(i)}\}_{g\in G}}$.
The order $\mathrm{ord}(\omega)$ of the anomaly is the smallest positive integer $n$ for which this diagonal symmetry is anomaly-free. 
It is equivalent to the order under stacking of the anomaly's inflow theory in one higher dimension.
If such a finite $n$ exists, the anomaly has a finite order ${\mathrm{ord}(\omega) = n}$. 
Otherwise, the anomaly has an infinite order, in which case we will say ${\mathrm{ord}(\omega) = \infty}$. 
In QFT, perturbative anomalies captured by Feynman diagrams, such as chiral anomalies, are of infinite order.
An anomaly is trivial, i.e., the symmetry is anomaly-free, if and only if ${\mathrm{ord}(\omega) = 1}$.

Infinite-order anomalies can also occur in quantum lattice systems. For example, they arise in Villain-type lattice models~\cite{SG190102637,GLS210301257,BCJ231017539,CS221112543,FS221113047,FXZ251017969,TPF260104304,S260114359,LSS260406307}, whose local degrees of freedom have infinite-dimensional Hilbert spaces and are subject to local Gauss law constraints. 
By contrast, infinite-order anomalies are much more elusive in lattice systems with finite-dimensional local Hilbert spaces, such as qubit systems.
Indeed, in ${1+1}$d lattice systems of this type, it has been proven that infinite-order anomalies of compact connected Lie group symmetries implemented by locality-preserving operators cannot be realized~\cite{KS240102533, L260213948}. 
A similar no-go theorem was proven in~\cite{FX230610105} for the chiral anomaly of a U(1) symmetry in ${3+1}$d.
These no-go theorems generalize the Nielsen-Ninomiya theorem~\cite{NN1981a, NN1981b, NN1981c, F1982}, which applies only to non-interacting fermions, to broader settings.

An anomaly in the UV implies that an anomaly arises in the IR.
When the UV and IR symmetry groups are the same, then their anomalies are also the same. The UV and IR anomalies, however, can differ when the UV and IR symmetry groups are different. In particular, the orders of UV and IR anomalies need not be the same.
As we show in \appref{sec:anomalyOrdRG}, the order of an anomaly can change under RG and satisfies
\begin{equation}\label{eq:rg_order_inequality}
    \mathrm{ord}(\omega_\mathrm{UV}) \leq \mathrm{ord}(\omega_\mathrm{IR}).
\end{equation}
The simplest scenario is when this inequality is saturated, and the anomalies match straightforwardly.
The scenario of an emergent anomaly corresponds to ${\mathrm{ord}(\omega_\mathrm{UV}) = 1}$ while ${\mathrm{ord}(\omega_\mathrm{IR}) > 1}$.
For example, an anomaly of an emergent symmetry is always emergent. (See \tabref{table} for a selection of more nontrivial examples.) Emergent anomalies, however, do not necessarily impose constraints on the UV theory.

In this paper, we present a new lattice anomaly in a quantum lattice system with finite-dimensional local Hilbert spaces and argue that, once an appropriate lattice CPT symmetry is imposed, this anomaly has infinite order. 
Our starting point is one of the simplest infinite-order anomalies in QFT: the chiral anomaly of a free massless Dirac fermion. 
In~\cite{CPS240912220}, it was shown that the vector and axial symmetries of the massless Dirac fermion in ${1+1}$d can arise in a lattice system with finite-dimensional local Hilbert spaces from a lattice symmetry governed by the Onsager algebra. 
The resulting Onsager symmetry has a lattice anomaly whose continuum limit is the chiral anomaly.

We show that the lattice anomaly of the Onsager symmetry alone is of finite order. 
In particular, it is an order two anomaly.
While this is distinct from the chiral anomaly in the continuum, we explain why there is no contradiction. 
We then show that imposing a lattice CPT symmetry resolves this mismatch. 
With the CPT symmetry enforced, the Onsager symmetry's lattice anomaly order is enhanced from two to infinite.
This additional symmetry is natural from the continuum perspective, since under the usual assumptions of the CPT theorem, it is present in every relativistic unitary local QFT.\footnote{Every unitary, Lorentz-invariant QFT has an antiunitary symmetry that simultaneously reflects one spatial coordinate and reverses time. It becomes the spacetime $\pi$ rotation in Euclidean signature. This symmetry is often referred to as the CRT symmetry \cite{W160502391}, with R for reflection. Most of the current paper is about ${1+1}$d systems, where there is no distinction between parity (P) and reflection (R), and we refer to this antiunitary symmetry as CPT. The only exception is \secref{sec:CRTSPT}, where we will discuss a CRT symmetry in ${2+1}$d, where R and P are different.}
This evades the aforementioned no-go theorems since (1) the Onsager symmetry group is an infinite-dimensional, non-compact Lie group and (2) the lattice CPT transformation is not locality-preserving. 
We further study the robustness of this anomaly under the addition of ancillas and describe the corresponding ${2+1}$d anomaly-inflow symmetry-protected topological (SPT) phase.

The remainder of this paper is as follows.
In \secref{sec:onsager_symmetries}, we review the Onsager algebra and its realization as a symmetry in fermionic quantum lattice systems. We also overview the locality properties and onsiteability of its symmetry operators.
In \secref{sec:lattice_anomalies}, we discuss the anomaly of this Onsager symmetry, showing that it is order two and clarifying why that is consistent with it becoming an infinite-order anomaly in the IR.
Then, in \secref{sec:lattice_crt}, we introduce the lattice CPT symmetry and show that it increases the order of this anomaly to infinity. 
In our discussion of anomalies in \secref{sec:lattice_anomalies} and \secref{sec:lattice_crt}, we further explore their robustness to adding ancillas as well as their corresponding ${2+1}$d anomaly inflow SPTs.
Throughout the main text, we give concise derivations of our results and present more technical details in the Appendix.

\section{Emergent anomalies in quantum mechanics}\label{anomQM}

We start with a simple example in quantum mechanics 
to illustrate the concept of emergent anomalies. 
(See \appref{QManomappsec} for a general model that realizes any emergent anomaly in quantum mechanics.)
In quantum mechanics, as opposed to higher-dimensional QFTs or lattice systems, a global symmetry $G$ has an anomaly if it is realized projectively on the Hilbert space. 
Inequivalent projective representations are classified by the group cohomology $H^2(G,\mathrm{U}(1))$.
Since there are no one-dimensional nontrivial projective representations, there are no $G$-invariant states, and an anomaly in quantum mechanics implies both an obstruction to gauging $G$ and an obstruction to realizing a nondegenerate ground state.

Consider a quantum-mechanical system of two qubits labeled by 1 and 2. Let $X_j$ and $Z_j$ with ${j=1,2}$ denote the Pauli matrices acting on the corresponding four-dimensional Hilbert space. 
We take the Hamiltonian of this four-level system to be
\begin{equation}\label{eq:two_qubit_hamiltonian}
    H = \left({E_1+E_2\over2}+{E_1-E_2\over2}X_1\right)\left(1+Z_2\over2\right),
\end{equation}
where the parameters ${E_1, E_2\in \mathbb{R}}$ satisfy ${E_2>E_1>0}$. 
This Hamiltonian has two ground states, both with zero energy, that span the ${Z_2=-1}$ eigenspace. 
The first and second excited states have energies $E_1$ and $E_2$, respectively, and span the ${Z_2=+1}$ eigenspace.

As we now show, the Hamiltonian \eqref{eq:two_qubit_hamiltonian} has an emergent anomaly.

The Hamiltonian \eqref{eq:two_qubit_hamiltonian} commutes with the  operators
\begin{equation}
A\equiv \mathsf{CZ},
\qquad
B\equiv X_1,~~~ C\equiv Z_2,
\end{equation}
where ${\mathsf{CZ} = \frac12(1+Z_1 +Z_2-Z_1Z_2)}$ is the controlled-Z gate. 
These symmetry operators satisfy
\begin{equation}\label{eq:abc_algebra}
A^2=B^2=C^2 =1,
\qquad
AB=CBA,
\end{equation}
and $C$ is central. 
Therefore, the symmetry group formed by $A$, $B$, and $C$ is ${G_\text{UV} = D_8}$,
the order 8 dihedral group. Here, "UV" indicates that this  is a global symmetry of the entire quantum mechanical system with the full four-dimensional Hilbert space. 
Moreover, from \eqref{eq:abc_algebra}, $A$, $B$, and $C$ furnish a linear representation of $D_8$, so the ${G_\text{UV} = D_8}$ symmetry is anomaly-free, i.e., ${\text{ord}(\omega_\text{UV})=1}$.

The "IR" of a quantum mechanical system corresponds to its ground state subspace. 
Since every ground state is in the ${C=Z_2=-1}$ eigenspace, $G_\text{UV}$ is not faithfully represented in the ground state subspace. 
Instead, the "IR" global symmetry $G_\text{IR}$ is a quotient of $G_\text{UV}$, generated by 
\begin{equation}
    A_\mathrm{IR} = A\,\Big|_{Z_2 = -1} \!\!= Z_1,
    \qquad
    B_\mathrm{IR} = B\,\Big|_{Z_2 = -1} \!\!= X_1.
\end{equation}
These IR symmetry operators satisfy
\begin{equation}
    A_\mathrm{IR}^2 = B_\mathrm{IR}^2 = 1,\qquad
    A_\mathrm{IR}B_\mathrm{IR} = -B_\mathrm{IR}A_\mathrm{IR}.
\end{equation}
Therefore, the ``IR'' global symmetry is a projective ${G_{\mathrm{IR}} = \mathbb Z_2\times\mathbb Z_2}$ symmetry, which is an anomalous symmetry.
Since the projective  phase is a sign (which is formally classified by ${H^2(\mathbb{Z}_2\times \mathbb{Z}_2,\mathrm{U}(1))=\mathbb{Z}_2}$), it is an order-2 anomaly, i.e., $\text{ord}(\omega_\text{IR})=2$.

This is a simple example of how an IR anomaly can emerge from an anomaly-free UV symmetry, as noted in \tabref{table}.
We note that while the other examples in \tabref{table} are in ${1+1}$d, they are qualitatively similar to this quantum-mechanical example: the faithful IR symmetry $G_\text{IR}$ is a quotient of the UV symmetry $G_\text{UV}$, and the order of the anomaly for $G_\text{IR}$ can be greater than that for $G_\text{UV}$.

\section{Onsager symmetries}
\label{sec:onsager_symmetries}

We start with a review of the Onsager symmetry and its lattice realization following \cite{CPS240912220}. 
An Onsager symmetry is a continuous symmetry whose conserved charges ${\{Q_n, G_m\}_{n,m\in\mathbb Z}}$ form the Onsager algebra~\cite{O1944}
\begin{subequations}\label{eq:onsager_algebra}
\begin{align}
[ Q_n,  Q_m] &= i   G_{m-n} \\
[ G_n,  G_m] &=0 \\
[ Q_n,  G_m] &= 2 i \left(  Q_{n-m} -  Q_{n+m }\right)
\end{align}
\end{subequations}
It is a non-abelian, infinite-dimensional Lie algebra. 
Note that ${G_n = -G_{-n}}$.
The Onsager algebra can be generated by $Q_0$ and $Q_1$. 
It also has infinitely many subalgebras that are isomorphic to the Onsager algebra.
For example, $Q_0$ and $Q_2$ also generate an Onsager algebra.\footnote{More generally, the charges $Q_{a}$ and $Q_{a+b}$ with fixed ${a\in\mathbb Z}$ and ${b\in \mathbb Z_{>0}}$ generate the algebra ${\{Q_{a+nb}, G_{mb}\}_{n,m\in\mathbb Z}}$, which is isomorphic to the Onsager algebra by ${Q_{a+nb} \mapsto Q_n}$ and ${G_{mb}\mapsto G_{m}}$. }

\subsection{Realization in fermionic systems}

Onsager symmetries can be realized in simple ${1+1}$d fermionic quantum lattice systems~\cite{VOF181209091, CPS240912220, PCS241218606, X250110837,Y250410263,AO260211696}.
Consider a finite lattice on a ring with $L$ sites and a complex fermion at each site $j$.
The complex fermion operators $c_j$ satisfy
the canonical anti-commutation relations ${\{c_j,c_{j'}\} =
 \{c_j^\dag, c_{j'}^\dag\} =0}$  and ${\{c_j , c_{j'}^\dag\} = \delta_{j j'}}$. 
We decompose $c_j$ as ${c_j =\frac{1}{2}( a_j + i b_j)}$, where the Majorana fermion operators ${a_j= a_j ^\dag}$ and ${b_j = b_j ^\dag}$ satisfy ${\{a_j, b_{j'}\}=0}$ and ${\{a_j, a_{j'}\}=\{b_j, b_{j'}\}= 2 \delta_{j j'}}$. 
Consider the Hermitian operators
\begin{align}\label{eq:fermion_onsager_charges} 
    Q_n = \frac{i}{2} \sum_{j=1}^L a_j b_{j+n}, \ \  
    G_n = \frac{i}{2} \sum_{j=1}^L a_j a_{j+n} - b_j b_{j+n}.
\end{align}
These operators obey the commutation relations~\eqref{eq:onsager_algebra}. Hence, in the ${L\to\infty}$ limit where $n$ and $m$ can take any value in $\mathbb Z$, they obey the Onsager algebra.

Each operator $Q_n$ has quantized eigenvalues.
First, note that the operator ${Q_0 = \sum_j\big(c_j^\dag c_j - \frac{1}{2}\big)}$  has eigenvalues $ \{ -\frac L2 , -\frac L2 +1, ... , \frac L2 \}$ since the eigenvalues of $c_j^\dag c_j$ are 0 or 1.
Then, to show this is true for a general $Q_n$, consider the Majorana translation operators $T_a$ and $T_b$ which satisfy
\begin{align}
T_a \begin{bmatrix}
    a_j\\
    b_j\\
\end{bmatrix} T_a^\dag = 
\begin{bmatrix}
    a_{j+1}\\
    b_{j}\\
\end{bmatrix},\qquad
T_b \begin{bmatrix}
    a_j\\
    b_j\\
\end{bmatrix} T_b^\dag = \begin{bmatrix}
    a_j\\
    b_{j+1}\\
\end{bmatrix}. 
\end{align}
Each $Q_n$ satisfies ${Q_n = T_b^n Q_0 T_b^{-n}}$. Therefore, since $T_b$ is unitary, each $Q_n$ has the same spectrum as $Q_0$. Throughout this paper, we will assume that $L$ is even, so all $Q_n$ have integer quantized eigenvalues.
Note that linear combinations of $Q_n$, such as ${\widetilde Q_n = \frac12(Q_{-n} + Q_n)}$ with ${n\neq 0}$, are not necessarily quantized.

The charge operators $Q_n$ and $G_n$ have corresponding unitary symmetry operators $e^{i\theta_n Q_n}$ and $e^{i\lambda_n G_n}$, respectively. 
Each $e^{i\theta_n Q_n}$ is a $\mathrm U(1)$ symmetry operator since the eigenvalues of $Q_n$ are quantized.
On the other hand, the eigenvalues of $G_n$ are not quantized, and each $e^{i\lambda_n G_n}$ is an $\mathbb{R}$ symmetry operator.

\subsection{Locality and onsiteability}

We now turn to the locality properties of these Onsager symmetry operators. 
The operator $e^{i\theta_n Q_n}$ acts on the Majorana fermions as
\begin{align}\label{eq:onsager_charge_adjoint_action}
e^{i \theta_n  Q_n} 
\begin{bmatrix}
    a_j\\
    b_j\\
\end{bmatrix}
e^{- i \theta_n Q_n}
&=
\begin{bmatrix}
   \cos (\theta_n)\,  a_j  + \sin (\theta_n)\,  b_{j+n}\\
   \cos (\theta_n)\, b_j - \sin (\theta_n) \, a_{j-n} \\
\end{bmatrix} \hspace{-0.15cm}.
\end{align}
Therefore, for all ${n =  O(L^0)}$, $e^{i\theta_n Q_n}$ is a strictly locality-preserving unitary, i.e., a quantum cellular automaton (QCA). In fact, $e^{i\theta_n Q_n}$ is a finite-depth quantum circuit (FDQC).
This can be seen using the index for fermionic QCAs.\footnote{Recall that the index of a QCA quantifies information transferred throughout space. The index of a one-dimensional QCA $U$ takes the form ${\mathrm{ind}_\mathrm{f}(U) = 2^{\zeta/2}\frac{p}{q}}$, where ${\zeta\in\{0,1\}}$ and $p,q$ are positive integers~\cite{FPP170307360}. When $U$ is a bosonic QCA, ${\zeta = 0}$ and $\mathrm{ind}_\mathrm{f}(U)$ becomes the GNVW index~\cite{GNV09103675}.} 
We denote the index of a QCA $U$ by $\mathrm{ind}_\mathrm{f}(U)$.
It satisfies ${\mathrm{ind}_\mathrm{f}(U) = 1}$ if and only if $U$ is an FDQC, and ${\mathrm{ind}_\mathrm{f}(U_1U_2) = \mathrm{ind}_\mathrm{f}(U_1)\mathrm{ind}_\mathrm{f}(U_2)}$.
Since $T_b$ is a QCA and
\begin{equation}\label{eq:majorana_translation_conjugation}
    e^{i\theta_n Q_n} = T_b^n e^{i\theta_n Q_0} T_b^{-n},
\end{equation}
the index of $e^{i\theta_n Q_n}$ is the same as that of $e^{i\theta_n Q_0}$. Therefore, because $e^{i\theta_n Q_0}$ is an FDQC, $e^{i\theta_n Q_n}$ is also an FDQC.

Each $e^{i\theta_n Q_n}$ with ${n= O(L^0)}$ can be made onsite using a QCA.\footnote{Relatedly, each U(1) symmetry represented by $e^{i\theta_n Q_n}$ is anomaly-free. Indeed, $Q_n$ itself is a valid local Hamiltonian that commutes with $e^{i\theta_n Q_n}$ and has a unique gapped ground state with eigenvalue $-L/2$. In fact, the $Q_0$ and $Q_1$ operators are the fixed-point Hamiltonians of the two invertible topological phases in the Kitaev chain \cite{K0506438}. } 
(If ${n= O(L)}$, then $Q_n$ is not locality-preserving and, thus, clearly cannot be made onsite using any locality-preserving unitary).
It follows from~\eqref{eq:majorana_translation_conjugation} that a disentangler for ${e^{i\theta_n Q_n}}$ is $T_b^{-n}$ since ${e^{i\theta_n Q_0} = \prod_{j=1}^L e^{- \frac{\theta_n}{2} a_j b_j}}$ is onsite. When $n$ is even and order $O(L^0)$, ${e^{i\theta_n Q_n}}$ can be disentangled by a FDQC. In particular, ${e^{i\theta_n Q_n}}$ for even $n$ satisfies
\begin{equation}
    (T_a T^{-1}_b)^{n/2} \,e^{i\theta_n Q_n} (T_a T^{-1}_b)^{-n/2} = e^{i\theta_n Q_0}.
\end{equation}
$(T_a T^{-1}_b)^{n/2}$ is an FDQC since its QCA index is 
\begin{equation}
    \mathrm{ind}_\mathrm{f}((T_a T^{-1}_b)^{n/2}) = \left(\sqrt{2}\frac1{\sqrt{2}}\right)^{n/2} = 1.
\end{equation}
(The QCA index of both $T_a$ and $T_b$ is $\sqrt{2}$~\cite{FPP170307360}).
When $n$ is odd, there is no FDQC that disentangles $e^{i\theta_n Q_n}$. The unitary $(T_a T^{-1}_b)^{(n-1)/2}$ is an FDQC that maps $Q_n$ to $Q_1$. 
Since the ground states of $Q_0$ and $Q_1$ as Hamiltonians are in different topological phases~\cite{K0506438}, there is no FDQC that maps $Q_n$ to $Q_0$~\cite[Prop 2]{HC14013820}. 
Since the only onsite operator, up to conjugating by an onsite unitary, with eigenvalues $\{-\frac{L}{2}, -\frac{L}{2} + 1, ..., \frac{L}{2}\}$ is $Q_0$, there is no FDQC that makes $Q_1$ onsite.

The unitary operator ${e^{i\sum_{n= O(L^{0})}(\theta_n Q_n + \lambda_n G_n)}}$ is not generally a QCA. However, it is a locality-preserving unitary when each ${\lambda_n = O(L^0)}$. Namely, it maps a local operator to a quasi-local operator (one with tails). This follows from the Lieb-Robinson bound~\cite{LR1972, BHV0603121}.
Indeed, ${\sum_{n= O(L^{0})}(\theta_n Q_n + \lambda_n G_n)}$ is a valid local Hamiltonian operator, i.e., it is Hermitian and a sum of strictly local operators.
Thus, the unitary ${e^{i\sum_{n= O(L^{0})}(\theta_n Q_n + \lambda_n G_n)}}$ is a valid time evolution operator and is locality-preserving when each ${\lambda_n = O(L^0)}$ by the Lieb-Robinson bound. 
This implies that in the thermodynamic limit, ${e^{i\sum_{n\in\mathbb{Z}}(\theta_n Q_n + \lambda_n G_n)}}$ with finitely many $\theta_n,\lambda_n$ nonzero is locality-preserving for all finite real $\lambda_n$.\footnote{More precisely, it is an automorphism of the algebra of quasi-local operators for the infinite-size quantum system.} Therefore, in the thermodynamic limit, $Q_0$ and $Q_1$ generate a locality-preserving symmetry.

\section{Finite order Onsager anomaly}
\label{sec:lattice_anomalies}

The Onsager symmetry generated by $Q_0$ and $Q_1$ is anomalous~\cite{CPS240912220}. In particular, every local symmetry Hamiltonian commuting with $Q_0$ and $Q_1$ is gapless. Indeed, the most general local Hamiltonian that commutes with both $Q_0$ and $Q_1$ is 
\begin{align}\label{eq:general_onsager_symmetric_hamiltonian} 
H = -\frac{i}{2} \sum_{n=1}^N \sum_{j=1}^L g_n \left(a_j a_{j+n} + b_j b_{j+n}\right), 
\end{align}
where $N$ is a positive integer independent of $L$ and ${g_n\in \mathbb{R}}$. This Hamiltonian is exactly solvable and easily verified to be gapless for all $g_n$. We note that there are, in fact, interacting terms that commute with $Q_0$ and $Q_1$, but they are extremely non-local and, hence, do not appear in~\eqref{eq:general_onsager_symmetric_hamiltonian}.

The Hamiltonian~\eqref{eq:general_onsager_symmetric_hamiltonian} with ${g_{n} = \delta_{1,n}}$ is called the staggered fermion model~\cite{BSK1976}. Its IR limit is described by a free massless Dirac fermion.
This QFT's global symmetries include the $\mathrm{U}(1)_\text{V}$ vector and $\mathrm{U}(1)_\text{A}$ axial symmetries. 
Together, they form a global ${[\mathrm{U}(1)_\text{V} \times  \mathrm{U}(1)_\text{A}]/\mathbb Z_2}$ symmetry, which has a mixed anomaly we refer to as the chiral anomaly~\cite{S1959,J1963}. 
The Onsager symmetry of~\eqref{eq:general_onsager_symmetric_hamiltonian} becomes the ${[\mathrm{U}(1)_\text{V} \times  \mathrm{U}(1)_\text{A}]/\mathbb Z_2}$ symmetry in the IR~\cite{CPS240912220}. In particular, $Q_0$ and $Q_1$ flow to the vector $\mathcal{Q}_\mathrm{V}$ and axial $\mathcal{Q}_\mathrm{A}$ charges, respectively. The chiral anomaly in the IR is realized in~\eqref{eq:general_onsager_symmetric_hamiltonian} by the Onsager symmetry's lattice anomaly.

The chiral anomaly of ${[\mathrm{U}(1)_\text{V} \times  \mathrm{U}(1)_\text{A}]/\mathbb Z_2}$ is an infinite-order anomaly. Indeed, its anomaly inflow theory is ${{\cal Z} = e^{\frac{i}{4\pi}(\mathrm{CS}(A_\mathrm{R}) - \mathrm{CS}(A_\mathrm{L}))}}$, where ${\mathrm{CS}(A) = \int AdA}$ is the U(1) Chern-Simons term while $A_\mathrm{R}$ and $A_\mathrm{L}$ are the background gauge fields for the U(1) symmetries generated by ${\mathcal{Q}_\mathrm{R} = \frac12(\mathcal{Q}_\mathrm{V} - \mathcal{Q}_\mathrm{A})}$ and ${\mathcal{Q}_\mathrm{L} = \frac12(\mathcal{Q}_\mathrm{V} + \mathcal{Q}_\mathrm{A})}$, respectively.
Since there is no finite $n$ satisfying ${{\cal Z}^n = 1}$, the anomaly is infinite-order. This then begs the question: what is the order of the Onsager symmetry's anomaly?
We now show that its anomaly has a finite order of two. Namely, while a single copy of the quantum lattice system does not admit a phase with a symmetric, unique gapped ground state, two copies of the system do.

Let us now show that the lattice anomaly is order two. Consider two copies of the original system, where now two species of complex fermions reside on each site on the lattice. We will denote the two fermion operators belonging to site $j$ by ${c_j^\uparrow = \frac12(a_j^\uparrow + i b_j^\uparrow)}$ and ${c_j^\downarrow = \frac12(a_j^\downarrow + i b_j^\downarrow)}$. (We use the $\uparrow$ and $\downarrow$ superscripts as a convenient notation, not to refer to physical spin-1/2 states.)
The diagonal Onsager symmetry of this copied system is generated by ${Q_0 \equiv Q_0^\uparrow + Q_0^\downarrow}$ and ${Q_1 \equiv Q_1^\uparrow + Q_1^\downarrow}$.
It is straightforward to check that the following Hamiltonian commutes with these charges and has a unique gapped ground state:
\begin{align}\label{eq:two_flavor_onsager_gapping_hamiltonian} 
H = \frac{ig}{2} \sum_{j=1}^L  
(a^ \uparrow_j a^ \downarrow_j + 
b^ \uparrow_j b^ \downarrow_j). 
\end{align}
Thus, the diagonal part of two copies of the Onsager symmetry is anomaly-free, which shows that the Onsager symmetry's anomaly is of order two.

The lattice anomaly being order two, i.e., $\mathrm{ord}(\omega_\mathrm{UV})=2$,  is entirely compatible with the IR chiral anomaly being infinite-order,  i.e., $\mathrm{ord}(\omega_\mathrm{IR})=\infty$. Let us show why, first explicitly and then more abstractly.

Consider two copies of the staggered fermion Hamiltonian deformed by \eqref{eq:two_flavor_onsager_gapping_hamiltonian}. 
While \eqref{eq:two_flavor_onsager_gapping_hamiltonian} corresponds to a mass term on the lattice, it does not in the IR.
Indeed, this lattice Hamiltonian has the single-particle dispersions  
\begin{align}\label{eq:two_flavor_staggered_dispersion} 
E_k^\pm  = 2\sin\left( \frac{2 \pi k }{L}\right) \pm g. 
\end{align}
For ${g>2}$, this spectrum is gapped. 
On the other hand, the continuum limit of \eqref{eq:two_flavor_onsager_gapping_hamiltonian} in this model is the chemical potential term ${i\mu  (\Psi_\text{L}^\uparrow )^\dag\Psi_\text{L}^\downarrow +i \mu(\Psi_\text{R}^\uparrow )^\dag\Psi_\text{R}^\downarrow  + \text{h.c.}}$, where $\Psi_{\text{L/R}}$ is the left/right-moving Weyl fermion. 
Note that this term is not invariant under the Lorentz or CPT symmetry transformations of the IR, and it need not be, as these are not UV symmetries.
The dispersion relation \eqref{eq:two_flavor_staggered_dispersion} becomes linear in momentum, and the chemical potential shifts the constant term.
No matter how large $\mu$ is, there is always a gapless momentum point, and the spectrum is never gapped. 
The chemical potential $\mu$ is proportional to $gL$.
Therefore, a finite chemical potential in the continuum limit corresponds to taking the limits ${L\to \infty}$ and ${g\to 0}$, and there is no contradiction.

More formally, the UV symmetries of the lattice system are related to those of the IR QFT by a group homomorphism ${\rho : G_\text{UV} \to G_\text{IR}}$.
Let us denote the group classifying the anomalies of a $G_\text{UV}$ (resp.\ $G_\text{IR}$) symmetry by $\mathrm{Anom}_{\mathrm{UV}}$ (resp.\ $\mathrm{Anom}_{\mathrm{IR}}$). In the simplest of cases, this group is given by group cohomology. 
The homomorphism $\rho$ induces a homomorphism ${\rho^\star : \mathrm{Anom}_{\mathrm{IR}} \to \mathrm{Anom}_{\mathrm{UV}}}$. Anomaly matching is the requirement that the IR anomaly $\omega_\mathrm{IR}\in \mathrm{Anom}_{\mathrm{IR}}$ must be related to the UV anomaly ${\omega_\mathrm{UV}\in \mathrm{Anom}_{\mathrm{UV}}}$ by ${\rho^\star (\omega_\mathrm{IR}) = \omega_\mathrm{UV}}$.
Thus, anomaly matching does not require the order of the UV anomaly to match that of the IR anomaly.
For the Onsager symmetry, its UV anomaly satisfies ${\omega_\mathrm{UV}^2 = 1}$ and generates the group ${\langle \omega_\mathrm{UV} \rangle = \mathbb Z_2}$, while the IR anomaly generates the group ${\langle \omega_\mathrm{IR} \rangle = \mathbb Z}$.
These are compatible given the homomorphism $\rho^\star$ satisfies ${\rho^\star (\mathbb Z) \cong \mathbb Z_2}$, i.e., ${\rho^\star(n) = n\bmod 2}$ for the IR anomaly ${\omega_{\mathrm{IR}}^n \in \langle \omega_\mathrm{IR} \rangle}$.

\subsection{Robustness to ancilla}\label{sec:robust}

Thus far, we have discussed the Onsager symmetry and its anomaly for lattice systems with a fixed local Hilbert space. We now explore the robustness of the anomaly under adding finite-dimensional ancillas.
Adding ancillas amounts to stacking the original system by a new system with the same symmetry, and then keeping the diagonal symmetry. 
If the original system's Hilbert space and $G$ symmetry operators are $\mathscr{H}$ and $\{U_g\}_{g\in G}$, adding ancillas with Hilbert space $\mathscr{H}^{\mathrm{anc}}$ and $G$ symmetry operators $\{U_g^{\mathrm{anc}}\}_{g\in G}$ modifies
\begin{equation}
    \begin{aligned}
        \mathscr{H} &\mapsto \mathscr{H} \otimes \mathscr{H}^{\mathrm{anc}},
        \\
        \{U_g\}_{g\in G} &\mapsto \{U_g\otimes U_g^{\mathrm{anc}}\}_{g\in G}.
    \end{aligned}
\end{equation}
It is often assumed that $U_g^{\mathrm{anc}}$ is an onsite symmetry operator.
However, since the Onsager algebra is infinite-dimensional, it has no finite-dimensional faithful representations.\footnote{A faithful, finite $n$-dimensional representation of a Lie algebra $\mathfrak{g}$ over $\mathbb{C}$ is an injective homomorphism ${\mathfrak{g}\hookrightarrow \mathfrak{gl}_n(\mathbb{C})}$. Injectivity requires ${\mathrm{dim}(\mathfrak{g})\leq \mathrm{dim}(\mathfrak{gl}_n(\mathbb{C})) = n^2}$. Therefore, an infinite-dimensional Lie algebra cannot have a faithful, finite-dimensional representation.} Thus, the quantum lattice systems and ancillas we consider do not admit onsite Onsager symmetry operators, and the ancilla's Onsager symmetry will necessarily be non-onsite.
Instead, as advocated in~\cite{PB260211266}, we will consider ancillas whose $G$ symmetry operators $\{U_g^{\mathrm{anc}}\}_{g\in G}$ are anomaly-free in the sense that they admit a gapped symmetric phase with a unique ground state.
The classification of anomaly-free Onsager symmetry operators is unknown, which prevents us from systematically exploring the anomaly's robustness. Instead, we will consider examples and verify that the anomaly is robust in those cases.

For instance, since the anomaly is order two, we can consider the ancillary system of two complex fermions per site with Onsager symmetry generated by ${Q_0 = Q_0^\uparrow + Q_0^ \downarrow}$ and ${Q_1 = Q_1^\uparrow + Q_1^ \downarrow}$. Indeed, as we show in \appref{sec:q_0_q_1_ancilla}, the original system remains gapless upon adding these ancillas.

A more nontrivial example of an anomaly-free Onsager symmetry in a system with a single complex fermion at each lattice site is one whose Onsager algebra is generated by $Q_0$ and $Q_2$.\footnote{We thank Arkya Chatterjee for discussions on this point.} 
This is anomaly-free since it commutes with the Hamiltonian 
\begin{align}\label{eq:q_zero_q_two_ancilla_hamiltonian} 
H = -\frac{i}{2} \sum_{\ell=1}^{L/2} (a_{2\ell} a_{2\ell+1}  + b_{2\ell} b_{2\ell+1}), 
\end{align}
which has a unique gapped ground state. 
Adding fermionic ancillas obeying this anomaly-free Onsager symmetry changes the Onsager charges to ${Q_0 = Q_0^\uparrow + Q_0^\downarrow}$ and ${Q_1 = Q_1^\uparrow + Q_2^\downarrow}$.
It is straightforward to show that every local Hamiltonian commuting with these charges is gapless. 
Indeed, as in \appref{sec:q_0_q_1_anomaly}, any symmetric Hamiltonian must also commute with a Majorana separating operator $e^{-\frac{i \pi }{2}Q_1} e^{i \frac{\pi}{2}Q_0}$ which acts as  
\begingroup
\renewcommand{\arraystretch}{1.3}
\begin{align}
e^{-\frac{i \pi }{2}Q_1} e^{i \frac{\pi}{2}Q_0}\begin{bmatrix}
    a_j^\uparrow\\
    b_j^\uparrow\\
    a_j^\downarrow\\
    b_j^\downarrow\\
 \end{bmatrix}e^{-i \frac{\pi}{2}Q_0} e^{\frac{i \pi }{2}Q_1} 
 &= 
 \begin{bmatrix}
    a_{j-1}^\uparrow\\
    b_{j+1}^\uparrow\\
    a_{j-2}^\downarrow\\
    b_{j+2}^\downarrow\\
 \end{bmatrix}. 
\end{align}
\endgroup
Along with $Q_0$, this enforces the Hamiltonian to be quadratic and contain no mixing between the two flavors or between $a$ and $b$. Therefore, the Hamiltonian for the first flavor must be of the form  \eqref{eq:general_onsager_symmetric_hamiltonian}, so the total Hamiltonian is gapless. 
Therefore, the anomaly of the Onsager symmetry generated by $Q_0$ and $Q_1$ is not trivialized by adding ancillas that transform under the anomaly-free Onsager symmetry generated by $Q_0$ and $Q_2$. 
More generally, the anomaly is robust under adding ancillas whose anomaly-free Onsager symmetry is generated by $Q_n$ and $Q_m$ with ${n=m~\bmod 2}$.

\subsection{${2+1}$d SPTs for the Onsager symmetry}\label{2+1dOnsSPTSec}

Anomalies of global symmetries are naturally related to SPTs in one dimension higher. 
For the Onsager symmetry, there are no symmetric product states and, therefore, no trivial SPTs.\footnote{To prove this, assume ${\bigotimes_{j=1}^L \ket{\psi_j}}$ is Onsager symmetric and, therefore, an eigenstate of $Q_0$ and $Q_1$. Being an eigenstate of $Q_0$ means each ${\ket{\psi_j} = \ket{n_j}}$, where ${c_j^\dag c_j\ket{n_j} = n_j \ket{n_j}}$. However, each ${\ket{\vec{n}} = \bigotimes_{j=1}^L \ket{n_j}}$ satisfies ${\bra{\vec{n}}Q_1\ket{\vec{n}} = 0}$ and ${Q_1\ket{\vec{n}} \neq 0}$ for all ${L>2}$. Thus, $\ket{\vec{n}}$ is never a $Q_1$ eigenstate, and there are no Onsager symmetric product states.} Therefore, when distinguishing two SPTs, we will consider interfaces between them rather than spatial boundaries. 
Moreover, we will show that the difference between these two SPTs is of order two.\footnote{The order of the difference between two SPT Hamiltonians $H_1$ and $H_2$ is the smallest integer $N_f$ such that there exists a symmetric, trivially gapped interface between $N_f$ copies of $H_1$ and $N_f$ copies of $H_2$. Naturally, if no such $N_f$ exists, the difference between the SPTs is said to be infinite-order.} 
See~\cite{FTA231209272,SS240401369,IO240815960,PSV251223706} for further discussions on interfaces between SPTs in the context of generalized symmetries.

Consider a ${2+1}$d system of even length $L_x$ and even width $L_y$ with a complex fermion at each site. We choose a representation of the Onsager algebra that is uniform in the $y$ direction 
\begin{align}\label{eq:bulk_onsager_charge} 
Q_n = \frac{i}{2}\sum_{x=1}^{L_x}\sum_{y=1}^{L_y } a_{x,y} b_{x+n,y}. 
\end{align}
Two trivially gapped symmetric Hamiltonians are
\begin{subequations}\label{eq:onsager_spt_hamiltonians}
\begin{align}
H_1 
&= \frac i2 \sum_{x=1}^{L_x}\sum_{\ell=1}^{L_y / 2} (a_{x,2\ell}a_{x,2\ell+1} + b_{x,2\ell}b_{x,2\ell+1}) \label{eq:onsager_spt_hamiltonian_even} \\
&=  i \sum_{x=1}^{L_x}\sum_{\ell=1}^{L_y / 2}
(c^\dag_{x,2\ell}c_{x,2\ell+1} -c^\dag_{x,2\ell+1}c_{x,2\ell}), \nonumber \\
H_2 
&= \frac i2 \sum_{x=1}^{L_x}\sum_{\ell=1}^{L_y / 2} (a_{x,2\ell-1}a_{x,2\ell} + b_{x,2\ell-1}b_{x,2\ell})\label{eq:onsager_spt_hamiltonian_odd}\\
&= i \sum_{x=1}^{L_x}\sum_{\ell=1}^{L_y / 2} 
(c^\dag_{x,2\ell-1}c_{x,2\ell} -c^\dag_{x,2\ell}c_{x,2\ell-1}). \nonumber
\end{align}
\end{subequations}
A symmetric interface between these two Hamiltonians always has an odd number of 
fixed-$y$ layers uncoupled to the 2+1d bulk (see \figref{fig:onsager_spt}). Therefore, by the arguments of the previous section, the interface between these two Hamiltonians is always gapless, and they represent distinct SPTs.
\newcommand{\siteVerticalSpacing}{0.5}
\pgfmathsetmacro{\siteCurveAngle}{atan(\siteVerticalSpacing*tan(70))}

\begin{figure}[t!]
\centering
\begin{tikzpicture}[scale=0.8]

\foreach \x in {0,0.75,1.5,2.75}{
    \draw[color=black!10] (\x,{-4*\siteVerticalSpacing})--(\x,{4*\siteVerticalSpacing});
}
    \filldraw[color=black!20, 
    ]
        (-0.15,{1.5*\siteVerticalSpacing})--
        (-0.15,{-1.5*\siteVerticalSpacing})--
        (2.9,{-1.5*\siteVerticalSpacing})--
        (2.9,{1.5*\siteVerticalSpacing})--
        cycle;
        \node[anchor=west, color=black!40] at (3.05,0) () {Interface};
\foreach \x in {0,0.75,1.5,2.75}{
    \node[] at (\x,{6.2*\siteVerticalSpacing}) () {$\vdots$};
    \node[] at (\x,{-5.6*\siteVerticalSpacing}) () {$\vdots$};

    \foreach \startbottom in {-5,-3}{
        \draw[very thick, color=MajRed]
        (\x,{\startbottom*\siteVerticalSpacing}) to[out=\siteCurveAngle, in={-\siteCurveAngle}]
        (\x,{(\startbottom+1)*\siteVerticalSpacing});

        \draw[very thick, color=MajBlue]
        (\x,{(\startbottom+1)*\siteVerticalSpacing}) to[out={-180+\siteCurveAngle}, in={180-\siteCurveAngle}]
        (\x,{\startbottom*\siteVerticalSpacing});
        }
    \foreach \starttop in {2,4}{
        \draw[very thick, color=MajRed]
        (\x,{\starttop*\siteVerticalSpacing}) to[out=\siteCurveAngle, in={-\siteCurveAngle}]
        (\x,{(\starttop+1)*\siteVerticalSpacing});

        \draw[very thick, color=MajBlue]
        (\x,{(\starttop+1)*\siteVerticalSpacing}) to[out={-180+\siteCurveAngle}, in={180-\siteCurveAngle}]
        (\x,{\starttop*\siteVerticalSpacing});
        }
    \foreach \y in {-4,-2,...,4}{
        \fill (\x,{\y*\siteVerticalSpacing}) circle (0.06cm);
        }
    \foreach \y in {-5,-3,...,5}{
        \filldraw[fill=white, draw=black] (\x,{\y*\siteVerticalSpacing}) circle (0.06cm);
        }
    }
    \node[anchor=east] at (-0.9,{0*\siteVerticalSpacing-0.05}) () {$y=$};
    \node[anchor=east] at (-0.5,{0*\siteVerticalSpacing}) () {$0$};
    \foreach \ylabel/\ynumber in {-4/4,-2/2}{
        \node[anchor=east] at (-0.8,{\ylabel*\siteVerticalSpacing}) () {$-$};
        \node[anchor=east] at (-0.5,{\ylabel*\siteVerticalSpacing}) () {$\ynumber$};
    }
    \foreach \ylabel in {2,4}{
        \node[anchor=east] at (-0.5,{\ylabel*\siteVerticalSpacing}) () {$\ylabel$};
    }

    \foreach \ylabel in {-4,0,4}{
        \node[] at ({(2.75+1.5)/2+0.05},{\ylabel*\siteVerticalSpacing}) () {$\cdots$};
    }
    \node[] at (3.55,{4*\siteVerticalSpacing}) () {$H_1$};
    \node[] at (3.55,{-4*\siteVerticalSpacing}) () {$H_2$};

    \draw[->] (0.75,{-6.6*\siteVerticalSpacing})--(1.75,{-6.6*\siteVerticalSpacing}) node[midway, anchor=north] () {$x$};
\end{tikzpicture}

\caption{An interface between Onsager SPT phases in 2+1d, with a three-layer interface region. Sites at even $y$ are shown in black and those at odd $y$ in white. Red and blue lines indicate a coupling between the $a$ and $b$ Majorana fermions, respectively, in the Hamiltonians \eqref{eq:onsager_spt_hamiltonians}.
An interface between $H_1$ and $H_2$ includes an odd number of fixed--$y$ layers. These layers may couple among themselves but they are uncoupled to the ${2+1}$d bulk.}
\label{fig:onsager_spt}
\end{figure}
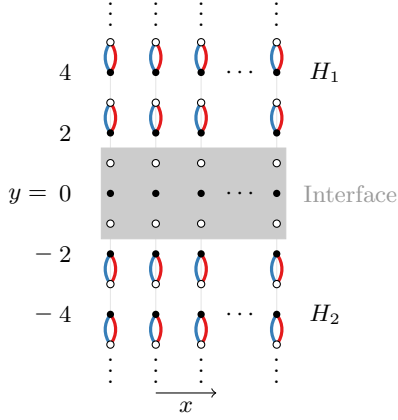

\subsection{An anomaly-free subgroup and the axial $\mathbb{R}$ symmetry}\label{U(1)xR section}

We have shown that the lattice anomaly of the Onsager symmetry is order two and, for the staggered fermion Hamiltonian, its continuum limit is the infinite-order chiral anomaly of ${[\mathrm{U}(1)_\mathrm{V}\times \mathrm{U}(1)_\mathrm{A}]/\mathbb Z_2}$.
Here, we point out that for the staggered fermion Hamiltonian, the same chiral anomaly in the continuum can also emerge from an anomaly-free subgroup of the lattice Onsager symmetry.

As observed in~\cite{CPS240912220}, there are other symmetries of the staggered fermion Hamiltonian that become the $\mathrm{U}(1)_\mathrm{A}$ symmetry in the continuum. For instance, the conserved charge operator ${\widetilde Q_1=\frac12 (Q_1+Q_{-1})}$ flows to the continuum axial charge~\cite{BSK1976, S1977}. 
However, $\widetilde Q_1$ is not quantized.
Together with $Q_0$, they generate a $\mathrm{U(1)}\times \mathbb{R}$ subgroup of the Onsager symmetry. 
This ${\mathrm{U(1)}\times \mathbb{R}}$ symmetry is anomaly-free.
To see this, we note that $Q_0$ itself is a Hamiltonian that commutes with $Q_0$ and $\widetilde Q_1$,
and as a Hamiltonian, it is 
gapped with a unique ground state. 
Therefore, the lattice symmetry ${\mathrm{U(1)}\times \mathbb{R}}$ is anomaly-free. 
This is an example of an infinite-order anomaly ($\mathrm{ord}(\omega_\mathrm{IR})=\infty$) emerging from an anomaly-free microscopic symmetry ($\mathrm{ord}(\omega_\mathrm{UV})=1$). 
We refer the reader to \appref{sec:spectral_flow} for further discussion.

\section{Infinite order anomaly from CPT}
\label{sec:lattice_crt}

Thus far, we have argued in \secref{sec:lattice_anomalies} that the lattice anomaly of the Onsager symmetry reviewed in \secref{sec:onsager_symmetries} is of order two.
Given that, in the staggered fermion Hamiltonian, this anomaly becomes the chiral anomaly in the IR, it is natural to wonder if its order can be enhanced from finite to infinite. Here, we will show that this can be done by imposing additional symmetries: namely, a lattice CPT symmetry.

In the lattice system with a single complex fermion on each site, consider the antiunitary operator $\Pi$ that satisfies
\begin{equation}\label{eq:lattice_cpt_fermion_action}
    \Pi\,i\,\Pi^{-1} = -i,\qquad
    \Pi \begin{bmatrix}
    a_j\\
    b_j\\
\end{bmatrix} \Pi^{-1} = \begin{bmatrix}
    a_{-j+1}\\
    b_{-j+1}\\
\end{bmatrix}.
\end{equation}
The general Onsager symmetric Hamiltonian \eqref{eq:general_onsager_symmetric_hamiltonian} commutes with $\Pi$.
Since $\Pi$ also satisfies ${\Pi^2=1}$~\cite[Eq~D.5]{SSZ250817115}, it generates an antiunitary $\mathbb Z^\mathrm{CPT}_2$ reflection symmetry.
Furthermore, it acts on the Onsager charges \eqref{eq:fermion_onsager_charges} as 
\begin{align}
\Pi Q_n \Pi^{-1} = -Q_{-n}, \qquad 
\Pi G_n \Pi^{-1} = -G_{-n}.
\end{align}
Therefore, the full symmetry group generated by ${(Q_0, Q_1, \Pi)}$ is ${\text{Onsager}\rtimes \mathbb Z_2^{\mathrm{CPT}}}$.
We will refer to $\Pi$ as a lattice CPT symmetry since it is an antiunitary reflection operator, satisfies $\Pi Q_0 \Pi^{-1} = -Q_{0}$, and becomes the CPT symmetry in the IR limit of the staggered fermion Hamiltonian~\cite{SSZ250817115}.

This ${\text{Onsager}\rtimes \mathbb Z_2^{\mathrm{CPT}}}$ symmetry has an anomaly since its Onsager symmetry is anomalous. To find the order of its anomaly, we consider $N_f$ copies of the system and enforce the diagonal Onsager and lattice CPT symmetries.
As shown in \appref{sec:q_0_q_1_ancilla}, the most general local Onsager symmetric Hamiltonian is 
\begin{align}\label{eq:diagonal_onsager_symmetric_hamiltonian}
\frac{i}{2} \sum_{n=-N}^N  \sum_{I,J=1}^{N_f}  \sum_{j=1}^L  g_n^{IJ}
(a^I_j a^J_{j+n} + 
b^I_j b^J_{j+n}).
\end{align}
Imposing the lattice CPT symmetry enforces the coupling constants $g_n^{IJ}$ to be an odd function of $n$.
This forbids mass terms like \eqref{eq:two_flavor_onsager_gapping_hamiltonian} and makes the dispersion relation of each term proportional to $\sin \left(\frac{2 \pi }{L} n k \right)$. Since the single particle dispersion is a polynomial in $\sin \left(\frac{2 \pi }{L} n k \right)$, it must be gapless at ${k=0}$. 
This is true for all finite values of $N_f$. 
Therefore, there is no $N_f$ for which the diagonal ${\text{Onsager}\rtimes \mathbb Z_2^{\mathrm{CPT}}}$ symmetry is anomaly-free, and the original ${\text{Onsager}\rtimes \mathbb Z_2^{\mathrm{CPT}}}$ symmetry's anomaly is infinite-order.

Note that the anomaly of $(Q_0, -Q_1, \Pi )$ forms an inverse of that in $(Q_0,Q_1, \Pi )$ in the sense that the diagonal subgroup of these two symmetries is non-anomalous. Explicitly, we take two complex fermions at each site $c_j^\uparrow$ and $c_j^\downarrow$ and impose the symmetries $Q_0 = Q_0^\uparrow + Q_0^\downarrow$ and $Q_1 = Q_1^\uparrow - Q_1^\downarrow$, 
as well as an antiunitary CPT symmetry $\Pi$ that acts on both flavors as in \eqref{eq:lattice_cpt_fermion_action}.
It is easy to check that the following Hamiltonian, originally written down in \cite[Eq 40]{X250110837}, is invariant under these symmetries
\begin{align}\label{eq:inverse_cpt_gapping_hamiltonian} 
H = \frac{ig}{2} \sum_{j=1}^L  (-1)^j
(a^ \uparrow_j a^ \downarrow_j + 
b^ \uparrow_j b^ \downarrow_j)
\end{align}
and has a unique gapped ground state.

\subsection{Robustness to ancilla}

Since a system of two complex fermions per site has an anomaly-free ${\mathrm{Onsager}\rtimes \mathbb Z^\mathrm{CPT}_2}$ symmetry generated by ${(Q^\uparrow_0+Q^\downarrow_0, Q_1^\uparrow-Q^\downarrow_1, \Pi)}$, we can use it as an ancilla.  
Indeed, as shown in \appref{sec:ancilla_q_0_minus_q_1}, the infinite-order anomaly of the ${(Q_0, Q_1, \Pi)}$ symmetry is robust to adding any number of these ancillas.

A more nontrivial ${\mathrm{Onsager}\rtimes \mathbb Z^\mathrm{CPT}_2}$ symmetric ancilla is the system with a single complex fermion per site whose Onsager symmetry is generated by $Q_0$ and $Q_2$. 
The ${\mathrm{Onsager}\rtimes \mathbb Z^\mathrm{CPT}_2}$ symmetry generated by ${(Q_0,Q_2,\Pi)}$ is anomaly-free because the trivially gapped Hamiltonian \eqref{eq:q_zero_q_two_ancilla_hamiltonian} is symmetric. 
We claim that the infinite-order anomaly of ${\mathrm{Onsager}\rtimes \mathbb Z_2^{\mathrm{CPT}}}$ is robust under stacking with this ancilla.
To see this, consider a system of ${N_f+1}$ complex fermions per site with an Onsager symmetry generated by ${Q_0 = \sum_{I=1}^{N_f+1}Q_0^{(I)}}$ and ${Q_1 = \sum_{I=1}^{N_f}Q_1^{(I)}
+ Q_2^{(N_f+1)}}$.
We regard the ${(N_f+1)}$th species as an ancilla. 
Following the argument in \secref{sec:robust}, the most general Onsager-symmetric Hamiltonian contains no couplings between the ancilla and the other $N_f$ fermions. 
Since the subsystem of the first $N_f$ fermions has been proven to be gapless for any integer $N_f$ once the $\mathbb Z_2^\mathrm{CPT}$ symmetry is imposed, it follows that the combined system is also gapless for all $N_f$ in the presence of the ancilla.

\subsection{${2+1}$d SPTs for the Onsager and CRT symmetry}\label{sec:CRTSPT}

Having discussed the infinite-order anomaly of the ${\mathrm{Onsager}\rtimes \mathbb Z_2^{\mathrm{CPT}}}$ symmetry in ${1+1}$d, we now present its corresponding anomaly-inflow theory, which is a ${2+1}$d SPT.
To do so, we first extend the ${1+1}$d ${\mathrm{Onsager}\rtimes \mathbb Z_2^{\mathrm{CPT}}}$ symmetry operators to ${2+1}$d. 
We extend the Onsager symmetry as in \secref{2+1dOnsSPTSec} and use the charge operators \eqref{eq:bulk_onsager_charge}.
The ${1+1}$d CPT symmetry in \eqref{eq:lattice_cpt_fermion_action} is extended to an antiunitary operator in the ${2+1}$d bulk which implements a link-centered reflection. 
We use the same symbol $\Pi$ to denote this ${2+1}$d operator and refer to it as the lattice CRT operator.
It is a CRT rather than a CPT operator in $2+1$d because it implements a reflection (R) transformation and not a parity (P) transformation. We take $\Pi$ to act on the fermions as\footnote{
This CRT symmetry can be made to act uniformly in the bulk with the following unitary transformation 
\begin{equation*}
    a_{x,y} \mapsto (-1)^{xy} a_{x,y},
    \qquad
    b_{x,y} \mapsto (-1)^{xy} b_{x,y}.
\end{equation*}
However, in this unitary frame, the SPT Hamiltonians \eqref{eq:onsager_spt_hamiltonians} and the Onsager charges are not uniform in the $y$ direction.
}
\begin{align}\label{eq:bulk_crt_fermion_action}
\Pi  \begin{bmatrix}
  a_{x,y}\\
 b_{x,y} \\
\end{bmatrix}\Pi^{-1}  = (-1)^y \begin{bmatrix}
  a_{-x+1, y}\\
  b_{-x+1,y}\\
\end{bmatrix}.
\end{align}
This commutes with the Onsager SPT Hamiltonians \eqref{eq:onsager_spt_hamiltonians} and forms ${\mathrm{Onsager}\rtimes \mathbb Z_2^{\mathrm{CRT}}}$ along with \eqref{eq:bulk_onsager_charge}. 
Therefore, the Hamiltonians \eqref{eq:onsager_spt_hamiltonians} are in distinct ${\mathrm{Onsager}\rtimes \mathbb Z_2^{\mathrm{CRT}}}$ SPT phases. At an interface of fixed $y$, the ${2+1}$d Onsager and CRT symmetries realize the ${\mathrm{Onsager}\rtimes \mathbb Z_2^{\mathrm{CPT}}}$ symmetry along the interface direction $x$. 

The ${\mathrm{Onsager}\rtimes \mathbb Z_2^{\mathrm{CPT}}}$ anomaly being infinite order implies that the difference between these ${\mathrm{Onsager}\rtimes \mathbb Z_2^{\mathrm{CRT}}}$ SPTs is also infinite order, which we now verify.
Because $H_1$ pairs even to odd $y$ Majoranas while $H_2$ pairs odd to even, every interface between $H_1$ and $H_2$ must have an odd number of fixed-$y$ layers that are uncoupled to the bulk $H_1$ and $H_2$ theories (see \figref{fig:onsager_spt}).
Therefore, the interface of the $N_f$-copied system can be viewed as the ${1+1}$d system considered in \appref{sec:ancilla_q_0_minus_q_1}, which cannot be trivially gapped.
Thus, the Hamiltonians \eqref{eq:onsager_spt_hamiltonians} represent distinct ${\mathrm{Onsager}\rtimes \mathbb Z_2^{\mathrm{CRT}}}$ SPTs whose difference is infinite-order.

\section{Conclusions}

The order of an anomaly is an important characterization of anomalous global symmetries, which cannot decrease under RG. 
In this paper, we explored how lattice CPT symmetry can affect lattice anomalies in ${1+1}$d. 
More precisely, we showed that the Onsager symmetry by itself has an order-two lattice anomaly, while the enlarged symmetry obtained by also imposing lattice CPT has an infinite-order anomaly. 
While infinite-order anomalies are ubiquitous in QFT, typical lattice anomalies, such as Lieb-Schultz-Mattis anomalies, have finite order. 
Our result provides the first example of an infinite-order anomaly in a lattice system with finite-dimensional local Hilbert spaces. It is also conceptually interesting: anomalies in quantum lattice systems are still not as systematically understood as their QFT counterparts, and the role of lattice CPT/CRT symmetry in this context remains largely unexplored.

The Onsager symmetry is a lattice realization of the chiral symmetry for a ${1+1}$d massless Dirac fermion, which has an infinite-order anomaly. 
It is therefore noteworthy that enforcing CPT symmetry on the lattice enhanced the lattice anomaly's order to that of the continuum. 
The Onsager symmetry is not an isolated example of this phenomenon, as illustrated in \tabref{table}.
Moreover, CPT---or more generally CRT---symmetry also plays an important role in the first-principles lattice theory of invertible phases of matter~\cite{O221209038, S250901711}. 
It is thus natural to ask what role CRT symmetry plays more generally in the relation between quantum lattice systems and relativistic QFTs. 
Further exploring this connection may provide a useful organizing principle for understanding which continuum anomalies admit faithful lattice counterparts.

Beyond the role of CRT symmetry, our results suggest two natural directions for future work.

First, it would be interesting to develop a more comprehensive understanding of the robustness of the Onsager symmetry's anomaly under the addition of ancillas. 
In particular, it would be interesting to study the anomalies of Onsager symmetries  (with or without CPT) realized on a qudit chain with ${d>2}$ \cite{VOF181209091} and their robustness, where free-fermion proofs for the anomalies no longer apply.   
More generally, it would be useful to identify an invariant index characterizing this anomaly that remains stable under the addition of anomaly-free ancillas.

A second direction is to study analogous phenomena in higher-dimensional lattice systems. 
This paper, as well as the examples summarized in \tabref{table}, focused on anomalies in ${1+1}$d. 
Lattice anomalies in higher dimensions, however, remain much less understood. 
(See~\cite{SZ260101191} for a ${2+1}$d example in which an order-eight lattice anomaly becomes order sixteen in the IR.)
For instance, there are Onsager-like symmetries in dimensions higher than ${1+1}$d that exhibit lattice anomalies~\cite{CPS250110862, GT250307708, PKC250504684, AKT251106198, KPS251204150, OY260326084}. 
It would be interesting to determine the orders of these anomalies, to understand whether CRT symmetry similarly refines or enhances them, and to clarify how their orders are related to those of the continuum QFTs.

\begin{acknowledgments}

We thank \"Omer Aksoy, Arkya Chatterjee, Christian Copetti, Lukasz Fidkowski, Dan Freed, Zohar Komargodski, and Max Metlitski for helpful discussions. 
We also thank Christian Copetti, Zohar Komargodski, and Wucheng Zhang for comments on a draft. 
E.L.S. is supported by the Hertz Foundation. 
S.D.P. and S.H.S. are  supported by the Simons Collaboration on Ultra-Quantum
Matter, which is a grant from the Simons Foundation
(651446, XGW; 651444, SHS).

\end{acknowledgments}

\appendix

\addtocontents{toc}{\protect\HideAppendixSubsectionsFromToc} 

\section{An RG inequality for anomalies}
\label{sec:anomalyOrdRG}

In this Appendix, we show that the order of an anomaly obeys the inequality~\eqref{eq:rg_order_inequality} between the UV and the IR:
\begin{equation}\label{eq:appendix_rg_order_inequality}
    \mathrm{ord}(\omega_\mathrm{UV}) \leq \mathrm{ord}(\omega_\mathrm{IR}).
\end{equation}

Let $G_\mathrm{UV}$ and $G_\mathrm{IR}$ be the UV and IR global symmetry groups, respectively.  
There is a group homomorphism $\rho: G_\mathrm{UV} \to G_\mathrm{IR}$, which is in general not surjective or injective. 
(See, for example, \cite{SS250508618} for more discussions of this map $\rho$.)  
We restrict our attention to symmetries whose anomalies are classified by groups (e.g., ordinary invertible 0-form symmetries). 
Let $\mathrm{Anom}_{\mathrm{UV}}$ and $\mathrm{Anom}_{\mathrm{IR}}$ denote the groups classifying the anomalies of the UV and IR symmetries, respectively. RG induces a group homomorphism 
\begin{equation}\label{eq:anomaly_pullback_map}
    \rho^\star : \mathrm{Anom}_{\mathrm{IR}} \to \mathrm{Anom}_{\mathrm{UV}}
\end{equation}
that relates the IR anomalies to the UV anomalies.
Suppose the IR anomaly is $\omega_\mathrm{IR}\in \mathrm{Anom}_{\mathrm{IR}}$. By anomaly matching, the UV anomaly ${\omega_\mathrm{UV}\in \mathrm{Anom}_{\mathrm{UV}}}$ is necessarily given by ${\omega_\mathrm{UV} = \rho^\star (\omega_\mathrm{IR})}$

The map~\eqref{eq:anomaly_pullback_map} can be justified from field theory, where  't Hooft anomalies are encoded by invertible topological field theories in one dimension higher. 
Specifically, the anomaly is characterized by a partition function $e^{i\alpha[A]}$, which assigns a phase to every spacetime manifold and background gauge field $A$. 
The set of anomalies clearly forms an abelian group: the product of two anomalies is $e^{i \alpha_1[A]+i\alpha_2[A]}$ and the inverse of $e^{i \alpha[A]}$ is $e^{- i \alpha[A]}$. 
Along an RG flow, every UV background gauge field $A_\mathrm{UV}$ of $G_\mathrm{UV}$ is mapped to an IR background configuration $A_\mathrm{IR}$, i.e., $A_\mathrm{IR}=  \Phi (A_\mathrm{UV})$ with the map $\Phi$ determined by $\rho$. 
This in turn induces the map \eqref{eq:anomaly_pullback_map} from the IR anomaly to the UV anomaly via 
$e^{ i \alpha_\mathrm{UV}[A_\mathrm{UV}]} = e^{i \alpha_\mathrm{IR} [\Phi(A_\mathrm{UV})]}$. 
We denote this map by $\rho^\star$ because in the case where the anomaly is classified by group cohomology, it denotes the pullback map of $\rho$. 
The map $\rho^\star$ is a homomorphism because $\rho$ is a homomorphism. 
Although no similarly systematic description of 't Hooft anomalies exists on the lattice, we expect an analogous map to $\rho^\star$ to exist as well.

Let us denote by $\operatorname{ord}(\omega_\text{IR})$ and $\operatorname{ord}(\omega_\text{UV})$ the order of the UV and IR anomalies, respectively.
There are four logically possible scenarios:
\begin{enumerate}
    \item \underline{$\omega_\mathrm{UV}$ and $\omega_\mathrm{IR}$ are finite-order:} From anomaly matching and the homomorphism properties of $\rho^\star$,
\begin{equation}\begin{aligned}\label{eq:uv_order_from_ir_order}
    \omega_\mathrm{UV}^{\mathrm{ord}(\omega_\mathrm{IR})} 
    &= \rho^\star (\omega_\mathrm{IR}^{\mathrm{ord}(\omega_\mathrm{IR})}) \\
     \implies \quad \omega_\mathrm{UV}^{\mathrm{ord}(\omega_\mathrm{IR})} 
    &= 1.
\end{aligned}\end{equation}
    This implies that ${\mathrm{ord}(\omega_\mathrm{UV}) \mid \mathrm{ord}(\omega_\mathrm{IR})}$ and, therefore, ${\mathrm{ord}(\omega_\mathrm{UV}) \leq  \mathrm{ord}(\omega_\mathrm{IR})}$. 
    \item \underline{$\omega_\mathrm{UV}$ is finite-order and $\omega_\mathrm{IR}$ is infinite-order:} From anomaly matching and the homomorphism properties of $\rho^\star$,
\begin{equation}\begin{aligned}
    \omega_\mathrm{UV}^{\mathrm{ord}(\omega_\mathrm{UV})} 
    &= \rho^\star (\omega_\mathrm{IR}^{\mathrm{ord}(\omega_\mathrm{UV})}) \\
    \implies \quad \rho^\star (\omega_\mathrm{IR}^{\mathrm{ord}(\omega_\mathrm{UV})})  &= 1.
\end{aligned}\end{equation}
    This is a constraint on $\omega_\mathrm{IR}$ that is compatible with $\omega_\mathrm{IR}$ being infinite-order. In this case, the order of the anomaly is enhanced from finite to infinite, and is consistent with~\eqref{eq:appendix_rg_order_inequality}.
    \item \underline{$\omega_\mathrm{UV}$ is infinite-order and $\omega_\mathrm{IR}$ is finite-order:} This is not possible. Indeed, from \eqref{eq:uv_order_from_ir_order}, if the IR anomaly is of finite order, then the UV anomaly is also of finite order.
    \item \underline{$\omega_\mathrm{UV}$ and $\omega_\mathrm{IR}$ are infinite-order:} This is possible. In this case, we say that the UV and IR anomalies have the same order, which is consistent with~\eqref{eq:appendix_rg_order_inequality}.
\end{enumerate}
Therefore, in all possible scenarios, the order of an anomaly must satisfy~\eqref{eq:appendix_rg_order_inequality}.

\section{Construction of emergent ${0+1}$d anomalies}\label{QManomappsec}

In this Appendix, we present a quantum mechanics model whose IR has an emanant $G$ symmetry with emergent anomaly ${[\omega]\in H^2(G,\mathrm{U}(1))}$. We will denote these anomalous $G$ symmetry operators by $\{U_g^\mathrm{IR}\}_{g\in G}$. They satisfy
\begin{equation}\label{eq:projective_ir_symmetry_algebra}
    U_g^\mathrm{IR} U_h^\mathrm{IR} = \omega(g,h)\, U_{gh}^\mathrm{IR},
\end{equation}
where ${\omega\colon G\times G\to\mathrm{U}(1)}$ is a normalized 2-cocycle with cohomology class $[\omega]$.

Consider the quantum mechanical system with Hilbert space ${\mathscr{H} = \mathrm{span}_\mathbb{C}\{\ket{0}, \ket{1}, \cdots, \ket{n}\}\cong \mathbb{C}^{n+1}}$ and Hamiltonian
\begin{equation}\label{eq:emergent_anomaly_hamiltonian}
    H = E\ketbra{n}{n}\qquad (E>0).
\end{equation}
This Hamiltonian has $n$ ground states with zero energy and a single excited state with energy $E$.
It commutes with the unitary operators
\begin{equation}
    U^{\mathrm{UV}}_g = U^{\mathrm{IR}}_g \oplus \mathbb{1}_1,
    \qquad
    \Omega^{\mathrm{UV}}_{g,h} = (\omega(g,h)\,\mathbb{1}_n) \oplus \mathbb{1}_1,
\end{equation}
where $\mathbb{1}_{m}$ denotes the $m \times m$ identity matrix.
These operators form an anomaly-free UV symmetry and satisfy
\begin{equation}\label{eq:nonabelian_uv_symmetry_algebra}
    U^{\mathrm{UV}}_g U^{\mathrm{UV}}_h = \Omega^{\mathrm{UV}}_{g,h}\, U^{\mathrm{UV}}_{gh}.
\end{equation}
They are anomaly-free since $-H$ is a symmetric Hamiltonian with a unique ground state.

The Hamiltonian \eqref{eq:emergent_anomaly_hamiltonian} has an emergent anomaly. In the ground state subspace, these UV symmetry operators become
\begin{equation}
    U^{\mathrm{UV}}_g \xrightarrow{~\mathrm{IR}~} U^{\mathrm{IR}}_g,
    \qquad
    \Omega^{\mathrm{UV}}_{g,h}
    \xrightarrow{~\mathrm{IR}~}  \omega(g,h)\,\mathbb{1}_n.
\end{equation}
In the IR, the nonabelian algebra \eqref{eq:nonabelian_uv_symmetry_algebra} becomes the projective $G$ algebra \eqref{eq:projective_ir_symmetry_algebra}. Therefore, the anomaly-free symmetry generated by $U^{\mathrm{UV}}_g$ and $\Omega^{\mathrm{UV}}_{g,h}$ becomes a $G$ symmetry in the IR. When the cohomology class $[\omega]$ is nontrivial, this $G$ symmetry is anomalous and \eqref{eq:emergent_anomaly_hamiltonian} has an emergent anomaly.

\section{Anomalies and ancillas of Onsager symmetry}
\label{sec:various_proofs}

In this Appendix, we will prove various statements about anomalies and ancillas of Onsager symmetries that are used in the main text.

\subsection{Review of the Onsager symmetry anomaly}
\label{sec:q_0_q_1_anomaly}

For completeness, we first review the proof in \cite[App B]{CPS240912220} that  \eqref{eq:general_onsager_symmetric_hamiltonian} is the most general local Hamiltonian commuting with 
\begin{align}
Q_0 = \frac{i}{2} \sum_{j=1}^L a_j b_j , \qquad 
Q_1 = \frac{i}{2} \sum_{j=1}^L a_j b_{j+1}. 
\end{align}
First, we define a Majorana separating operator as $M = e^{-\frac{i \pi }{2}Q_1} e^{i \frac{\pi}{2}Q_0}$ which, from  \eqref{eq:onsager_charge_adjoint_action}, acts on the fermions as
\begin{equation}\begin{aligned}\label{eq:majorana_separating_action} 
&e^{-\frac{i \pi }{2}Q_1} e^{i \frac{\pi}{2}Q_0}
\begin{bmatrix}
    a_j \\
    b_j\\
\end{bmatrix}
e^{-i \frac{\pi}{2}Q_0} 
e^{\frac{i \pi }{2}Q_1}
= \begin{bmatrix}
    a_{j-1}\\
    b_{j+1}\\
\end{bmatrix} .
\end{aligned}\end{equation}
No local, $M$-symmetric Hamiltonian can contain terms that mix $a$ and $b$. This occurs since one can apply $M$ order $L$ times, and the resulting symmetry-transformed Hamiltonian would be non-local if it mixes $a$ and $b$. Additionally, enforcing commutation with $Q_0$ prevents local interacting terms from appearing in the Hamiltonian.
The most general non-interacting local Hamiltonian that does not mix $a$'s and $b$'s is of the form 
\begin{align}
H = -\frac{i}{2} \sum_{n=1}^N \sum_{j=1}^L g_n(j) a_j a_{j+n} + \hat g_n(j) b_j b_{j+n}. 
\end{align}
This Hamiltonian will commute with $Q_0$ if and only if $g_n(j) = \hat g_n(j)$. It will commute with $M$ only if $g_n(j)$ are independent of $j$. It is a direct check that this Hamiltonian also commutes with $Q_1$, which shows that \eqref{eq:general_onsager_symmetric_hamiltonian} is the most general Onsager-symmetric local Hamiltonian. 

The Hamiltonian \eqref{eq:general_onsager_symmetric_hamiltonian} never has a unique gapped ground state.
In terms of complex fermions, \eqref{eq:general_onsager_symmetric_hamiltonian} reads
\begin{equation}
\begin{aligned}
H 
&= -i\sum_{n=1}^N \sum_{j=1}^L 
g_n\left(c_j^\dag c_{j+n} - c_{j+n}^\dag c_j\right)\\
&= 2\sum_{n=1}^N \sum_{k=1}^L 
g_n  \sin \left(\frac{2 \pi }{L} nk\right) 
\gamma_k^\dag \gamma_k .
\end{aligned}
\label{eq:general_onsager_hamiltonian_momentum_space}
\end{equation}
where ${\gamma_k = \frac{1}{\sqrt{L} } \sum_{j=1}^L  e^{- \frac{2 \pi i }{L}jk} c_j}$ satisfy ${\{\gamma_k, \gamma_{k'} ^\dag\} = \delta_{k,k'}}$. \eqref{eq:general_onsager_hamiltonian_momentum_space} is always gapless at ${k=0}$ if any $g_n$ are non-vanishing and, thus, the Onsager symmetry generated by $Q_0$ and $Q_1$ is anomalous.

\subsection{Onsager symmetric ancillas}
\label{sec:q_0_q_1_ancilla}

In this section, we show that the anomaly of the Onsager symmetry generated by $Q_0$ and $Q_1$ is robust to adding ancillas. We consider ancillas comprised of two complex fermions per site with an anomaly-free Onsager symmetry generated by ${Q_0^\uparrow + Q_0^\downarrow}$ and ${Q_1^\uparrow + Q_1^\downarrow}$. This robustness is equivalent to the following: in a system of $N_f$ complex fermions per site, the Onsager symmetry generated by 
\begin{align}\label{eq:flavor_diagonal_onsager_charges}
Q_0 = \sum_{I=1}^{N_f} Q_0^I , \quad 
Q_1 = \sum_{I=1}^{N_f}  Q_1^I \,,
\end{align}
is anomalous when $N_f$ is odd.

Since the $Q_n^I$ and $Q_n^{I'}$ commute for $I \ne I'$, the operator \eqref{eq:majorana_separating_action} acts on the fermions as 
\begingroup
\renewcommand{\arraystretch}{1.3}
\begin{align}
M
\begin{bmatrix}
    a_j^I \\
    b_j^I\\
\end{bmatrix} M ^{-1}
= \begin{bmatrix}
    a_{j-1}^I\\
    b_{j+1}^I\\
\end{bmatrix}\,. 
\end{align}
\endgroup
An argument entirely analogous to that following \eqref{eq:majorana_separating_action} then shows that local, $M$-symmetric Hamiltonians are quadratic and contain no mixing between $a$ and $b$. 
Imposing that the most general non-interacting Hamiltonian commutes with $Q_0$ and $M$ restricts it to the form 
\begin{align}
H
&= -\frac{i}{2}\sum_{I,J=1}^{N_f}  \sum_{n=-N}^N \sum_{j=1}^L  
g_n^{IJ} \left( a^{ I }_j a^{ J}_{j+n} +    b^{ I }_j b^{ J}_{j+n} \right) \,,
\end{align} 
where $g_n^{IJ} \in \mathbb R$ from Hermiticity and ${g_n^{IJ} = -g_{-n}^{JI}}$ from resumming ${j \to j-n}$. In terms of complex fermions in momentum space, this becomes 
\begin{align}\label{eq:flavor_onsager_momentum_hamiltonian}
H =  \sum_{I,J}\sum_{k}
h^{IJ}_k
(\gamma_k^{I})^\dag \gamma_k^J , \quad
h^{IJ}_k = -2 i \sum_{n=-N}^N e^{\frac{2 \pi i }{L} nk} g^{IJ}_n \,.
\end{align}
The coefficient $h^{IJ}_k $ is a Hermitian matrix, so it may be diagonalized.
The $N_f$ one-particle energies of $H$ are the eigenvalues of $h^{IJ}_k$. 
Evaluated at zero momentum, ${h^{IJ}_0 = -2i  \sum_{n=-N}^N g^{IJ}_{n}}$ is purely imaginary and anti-symmetric. Eigenvalues of such a matrix come in pairs $\pm \lambda $. 
When $N_f$ is odd, $h$ is odd-dimensional, and there must be at least one zero eigenvalue. Then, the one-particle spectrum has at least one zero eigenvalue at ${k=0}$, and since $h_k^{IJ}$ is a smooth function of the momenta $k$, we conclude $H$ is gapless at ${k=0}$.

\subsection{The infinite-order anomaly of ${\mathrm{Onsager}\rtimes\mathbb Z_2^\mathrm{CPT}}$}
\label{sec:q_0_q_1_crt_anomaly}

Here, we consider the same setup as the previous section, but now impose an additional CPT symmetry that acts as \eqref{eq:lattice_cpt_fermion_action}. 
As shown in the previous section, the most general Onsager symmetric Hamiltonian is \eqref{eq:flavor_onsager_momentum_hamiltonian}. Imposing CPT enforces that $g_n^{IJ} =  g_{n}^{JI} $ which means $g_n^{IJ} = -g_{-n}^{IJ}$. Thus, 
\begin{align}
H &= \sum_{I,J=1}^{N_f}  \sum_{k=1}^{L} h^{IJ}_k 
  (\gamma_k^I)^\dag\gamma_k^J,\nonumber\\
 h^{IJ}_k &= 4 \sum_{n=1}^N 
g_{n}^{IJ} \sin \left(\frac{2 \pi }{L}  n k\right).
\label{eq:cpt_flavor_onsager_momentum_hamiltonian}
\end{align}
Since every entry of $h^{IJ}_k$ is a linear combination of $\sin \left(\frac{2 \pi }{L} nk\right)$ its determinant must vanish smoothly at $k=0$. Therefore, at least one eigenvalue must vanish smoothly at $k=0$. We conclude that the Hamiltonian \eqref{eq:cpt_flavor_onsager_momentum_hamiltonian} is gapless for all $N_f$ and, thus, the anomaly of ${\mathrm{Onsager}\rtimes\mathbb Z_2^\mathrm{CPT}}$ is infinite order.

\subsection{${\mathrm{Onsager}\rtimes\mathbb Z_2^\mathrm{CPT}}$ symmetric ancillas}\label{sec:ancilla_q_0_minus_q_1}

In the main text, we showed that the Onsager symmetry generated by ${(Q^\uparrow_0+Q^\downarrow_0, Q_1^\uparrow+Q^\downarrow_1)}$ is anomaly-free due to the existence of the Hamiltonian \eqref{eq:two_flavor_onsager_gapping_hamiltonian} which has a unique gapped ground state. 
We further showed that the anomaly of the Onsager symmetry generated by ${(Q_0, Q_1)}$ persists under stacking with these anomaly-free ancilla. 
Here, we will extend this discussion to include ancillas with CPT symmetry.
In particular, we will consider the ancilla CPT symmetry to be generated by the antiunitary operator $\Pi$ that satisfies
\begingroup
\renewcommand{\arraystretch}{1.3}
\begin{align}\label{eq:anomaly_free_ancilla_cpt_action} 
\Pi \begin{bmatrix}
    a_j^\uparrow\\
    b_j^\uparrow\\
    a_j^\downarrow\\
    b_j^\downarrow\\
 \end{bmatrix}
 \Pi  ^{-1}
 &= 
 \begin{bmatrix}
    a_{-j+1}^\uparrow\\
    b_{-j+1}^\uparrow\\
    -a_{-j+1}^\downarrow\\
    -b_{-j+1}^\downarrow\\
 \end{bmatrix}. 
\end{align}
\endgroup
We now show that the order of the ${\mathrm{Onsager}\rtimes \mathbb Z^\mathrm{CPT}_2}$ anomaly persists under stacking with these ancillas. 

To begin, consider the more general system of $N_f$ complex fermions per site and the ${\mathrm{Onsager}\rtimes \mathbb Z^\mathrm{CPT}_2}$ symmetry generated by the diagonal Onsager charges~\eqref{eq:flavor_diagonal_onsager_charges}
and 
an antiunitary operator $\Pi$ that acts as
\begingroup
\renewcommand{\arraystretch}{1.3}
    \begin{equation}\begin{aligned}\label{eq:flavor_dependent_cpt_action} 
    &\Pi \begin{bmatrix}
    a_j^I\\
    b_j^I\\
\end{bmatrix} \Pi  ^{-1} 
= s_I \begin{bmatrix}
    a_{-j+1}^I\\
    b_{-j+1}^I\\
\end{bmatrix}\,,
\qquad \Pi\, i\, \Pi^{-1} = -i, \nonumber\\
&s_I =
\begin{cases}
1, & 1\le I\leq N_+\\
-1, & N_+ < I \le N_f
\end{cases} ,
    \end{aligned}\end{equation}
\endgroup 
with some integer $N_+$ obeying $0\le N_+ \le N_f$. 
This ${\mathrm{Onsager}\rtimes \mathbb Z^\mathrm{CPT}_2}$ symmetry is anomalous as long as ${N_+ \ne \frac{N_f}{2}}$. To see this, first recall that the most general Hamiltonian commuting with the 
diagonal Onsager symmetry \eqref{eq:flavor_diagonal_onsager_charges} is \eqref{eq:flavor_onsager_momentum_hamiltonian}. Imposing the CPT symmetry \eqref{eq:flavor_dependent_cpt_action} enforces that ${g_{n}^{IJ} = -s_I s_J g_{-n}^{IJ}}$. The matrix $h_k^{IJ}$ at zero momentum takes a particularly simple form:
\begin{align} 
 h^{IJ}_0 = - i (1 - s_I s_J)\sum_{n=-N}^N 
g_{n}^{IJ}
 = \begin{bmatrix}
    0 & A\\
    A^\dag & 0\\
 \end{bmatrix}, 
\end{align}
where $A$ is an ${N_+ \times (N_f-N_+)}$ matrix.  
If $N_+ \ne \frac{N_f}2$ then $A$ is a non-square rectangular matrix and either $A$ or $A^\dag$ must have more columns than rows. Therefore, either $A$ or $A^\dag$ must have a non-trivial kernel, which in turn gives rise to a zero eigenvector for $h_0^{IJ}$.  Hence, any Hamiltonian with $N_+ \ne \frac{N_f}2$ must be gapless.  
This implies that any number of ancillas with ${\mathrm{Onsager}\rtimes \mathbb Z^\mathrm{CPT}_2}$ symmetry generated by ${(Q^\uparrow_0+Q^\downarrow_0, Q_1^\uparrow+Q^\downarrow_1)}$ and \eqref{eq:anomaly_free_ancilla_cpt_action} stacked with $N_\text{copy} \geq 1 $ copies of the anomalous ${\mathrm{Onsager}\rtimes \mathbb Z^\mathrm{CPT}_2}$ symmetry is anomalous since $N_+ = \frac{N_f}{2} + N_\text{copy} \ne \frac{N_f}2$.

\section{Some fermionic crystalline SPTs}
\label{sec:maj_translations_spt}
In the main text, we discussed ${2+1}$d fermionic SPTs for the Onsager symmetry. Here, we review fermionic invertible phases in $1+1$d and SPTs for Majorana translations in 2+1d.

\subsection{Review of ${1+1}$d fermionic invertible phases}

\begin{figure}[t]
\centering
\begin{tikzpicture}
\def\nnodes{16}
\pgfmathtruncatemacro{\oddlast}{\nnodes/2-1}
\pgfmathtruncatemacro{\evenfirst}{\nnodes/2+2}
\pgfmathtruncatemacro{\evennext}{\nnodes/2+4}
\pgfmathtruncatemacro{\evenlast}{\nnodes-2}

\draw[samples=100,domain=0:360, shift={(0,0)}, color=black!10] plot ({1*cos(\x)},{1*sin(\x)});

\foreach \h in {1,...,\nnodes}{
  \pgfmathsetmacro{\angle}{90 - 360*\h/\nnodes}
  \node[inner sep=0] at (\angle:1) (n\h) {};
  }
\foreach \x in {1,3,...,\oddlast}{
  \pgfmathtruncatemacro{\y}{mod(\x,\nnodes)+1}
  \pgfmathsetmacro{\startangle}{90 - 360*\x/\nnodes}
  \pgfmathsetmacro{\endangle}{90 - 360*\y/\nnodes}
  \draw[ultra thick, color=MajBlue] (n\x)
    arc[start angle=\startangle, end angle=\endangle, radius=1];
}

\foreach \x in {\evenfirst,\evennext,...,\evenlast}{
  \pgfmathtruncatemacro{\y}{mod(\x,\nnodes)+1}
  \pgfmathsetmacro{\startangle}{90 - 360*\x/\nnodes}
  \pgfmathsetmacro{\endangle}{90 - 360*\y/\nnodes}
  \draw[ultra thick, color=MajBlue] (n\x)
    arc[start angle=\startangle, end angle=\endangle, radius=1];
}

\foreach \h in {2,4,...,\nnodes}{
  \pgfmathsetmacro{\angle}{90 - 360*\h/\nnodes}
  \fill (\angle:1) node[inner sep=0] (n\h) {} circle(0.04);
  }
\foreach \h in {1,3,...,\nnodes}{
  \pgfmathsetmacro{\angle}{90 - 360*\h/\nnodes}
  \filldraw[fill=white, draw=black] (\angle:1) node[inner sep=0] (n\h) {} circle(0.04);
  }

\node[anchor=south] at (90:1) {$j=0$};
\node[anchor=north] at (-90:1) {$j=\frac L2$};

\node[anchor=west] at (0:1) () { $H_2$};
\node[anchor=east] at (180:1) () { $H_1$};
\end{tikzpicture}  
\qquad
\begin{tikzpicture}
\def\nnodes{14}
\pgfmathtruncatemacro{\oddlast}{\nnodes/2-2}
\pgfmathtruncatemacro{\evenfirst}{\nnodes/2+1}
\pgfmathtruncatemacro{\evennext}{\nnodes/2+3}
\pgfmathtruncatemacro{\evenlast}{\nnodes-2}
\draw[samples=100,domain=0:360, shift={(0,0)}, color=black!10] plot ({1*cos(\x)},{1*sin(\x)}); 

\foreach \h in {1,...,\nnodes}{
  \pgfmathsetmacro{\angle}{90 - 360*\h/\nnodes}
  \node[inner sep=0] at (\angle:1) (n\h) {};
  } 
\foreach \x in {1,3,...,\oddlast}{
  \pgfmathtruncatemacro{\y}{mod(\x,\nnodes)+1}
  \pgfmathsetmacro{\startangle}{90 - 360*\x/\nnodes}
  \pgfmathsetmacro{\endangle}{90 - 360*\y/\nnodes}
  \draw[ultra thick, color=MajBlue] (n\x)
    arc[start angle=\startangle, end angle=\endangle, radius=1];
}

\foreach \x in {\evenfirst,\evennext,...,\evenlast}{
  \pgfmathtruncatemacro{\y}{mod(\x,\nnodes)+1}
  \pgfmathsetmacro{\startangle}{90 - 360*\x/\nnodes}
  \pgfmathsetmacro{\endangle}{90 - 360*\y/\nnodes}
  \draw[ultra thick, color=MajBlue] (n\x)
    arc[start angle=\startangle, end angle=\endangle, radius=1];
}

\foreach \h in {2,4,...,\nnodes}{
  \pgfmathsetmacro{\angle}{90 - 360*\h/\nnodes}
  \fill (\angle:1) node[inner sep=0] (n\h) {} circle(0.04);
  } 
\foreach \h in {1,3,...,\nnodes}{
  \pgfmathsetmacro{\angle}{90 - 360*\h/\nnodes}
  \filldraw[fill=white, draw=black] (\angle:1) node[inner sep=0] (n\h) {} circle(0.04);
  } 

\node[anchor=south] at (90:1) {$j=0$};
\node[anchor=north] at (-90:1) {$j=\frac L2$};

\node[anchor=west] at (0:1) () { $H_2$};
\node[anchor=east] at (180:1) () { $H_1$};
\end{tikzpicture}
\caption{Colored lines indicate a coupling between the Majorana fermions at those sites in the example interface Hamiltonians \eqref{eq:majorana_chain_interface_hamiltonian}. Even sites are colored in black, while odd sites are colored in white. The index $j$ runs from $0$ to ${L-1}$ going clockwise.
(Left) For ${L \equiv 0\bmod 4}$, there are unpaired Majorana fermions at ${j=0}$ and ${j=\frac L2 +1}$.
(Right) For ${L \equiv 2\bmod 4}$, there are unpaired Majorana fermions at ${j=0}$ and ${j=\frac L2}$. 
} 
\label{fig:spt_coupling}
\end{figure}
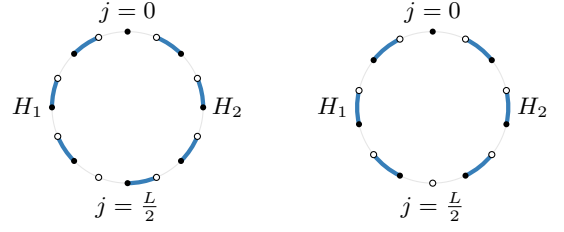

Consider a ${1+1}$d lattice system of $L$ sites where a single Majorana fermion $\chi_j$ resides at each site $j$, $L$ is even, and there are periodic boundary conditions. 
There are two invertible fermionic phases of this system. 
The Hamiltonians 
\begin{align}\label{eq:invertible_majorana_chain_hamiltonians} 
&H_1 = i \sum_{\ell=1}^{L/2} \chi_{2\ell} \chi_{2\ell+1} , \qquad H_2  = i \sum_{\ell=1}^{L/2} \chi_{2\ell-1} \chi_{2\ell}                       \,,
\end{align}
are exactly solvable gapped Hamiltonians that respectively realize these distinct phases~\cite{K0010440}. One way to see that $H_1$ and $H_2$ are in distinct invertible phases is that there are no trivial interfaces between these two theories. 
Consider the theory with interfaces between $H_1$ and $H_2$ at sites ${j=0}$ and ${j=L/2}$ that is described by the Hamiltonian
\begin{align}\label{eq:majorana_chain_interface_hamiltonian}
& \ \ H_\mathrm{interface} 
= \\
&\begin{dcases}
i \sum_{\ell=1}^{L/4} \chi_{2\ell-1} \chi_{2\ell}+
i \sum_{\ell=L/4+1}^{L/2-1} \chi_{2\ell} \chi_{2\ell+1}  ~~~ L = 0 ~ \operatorname{mod} ~ 4 \\ 
i \sum_{\ell=1}^{\frac{L-2}4} \chi_{2\ell-1} \chi_{2\ell}+
i \sum_{\ell=\frac{L-2}4+1}^{L/2-1} \chi_{2\ell} \chi_{2\ell+1}~~~ L = 2 ~\operatorname{mod}  ~ 4 
\end{dcases} \nonumber 
\end{align}
A graphical representation of this interface Hamiltonian is shown in \figref{fig:spt_coupling}. It is straightforward to show that $H_\mathrm{interface}$ is gapped and has two degenerate ground states. Indeed, there is an unpaired Majorana fermion near each interface.
The two-fold degeneracy is robust to deforming $H_\mathrm{interface}$ near the interfaces, and is protected by fermion number parity symmetry.
Hence, $H_1$ and $H_2$ are in different fermionic invertible phases.

Since ${1+1}$d invertible fermionic topological field theories are $\mathbb Z_2$ classified~\cite{KTT14067329}, two copies of $H_1$ should be in the same phase as two copies of $H_2$. This is, indeed, true. Consider the two-copied system with Majorana fermions $\chi^\uparrow_j$ and $\chi^\downarrow_j$ at each site and interface Hamiltonian ${H^\uparrow_\mathrm{interface} + H^\downarrow_\mathrm{interface}}$. The degenerate interface modes of this Hamiltonian can be lifted using a simple fermion number symmetric deformation. When ${L=0\bmod 4}$, this deformation is ${i( \chi^\uparrow_0\chi^\downarrow_0 +  \chi^\uparrow_{L/2+1}\chi^\downarrow_{L/2+1})}$. When ${L=2\bmod 4}$, this deformation is ${i( \chi^\uparrow_0\chi^\downarrow_0 + \chi^\uparrow_{L/2}\chi^\downarrow_{L/2})}$. Thus, two copies of $H_1$ is in the same phase as two copies of $H_2$.

\subsection{${2+1}$d SPTs with Majorana translations}\label{2dMajTranlSPT}

Here, we discuss ${2+1}$d fermionic SPTs protected by one-dimensional Majorana translation symmetry. 
Consider a two-dimensional square lattice of ${L_x\times L_y}$ sites with a Majorana fermion $\chi_{x,y}$ at each lattice site ${(x,y)}$.
We assume that $L_y$ is even and consider periodic boundary conditions in the $x$ and $y$ directions.
Suppose the corresponding lattice system has a Majorana translation symmetry generated by the unitary $T_\chi$ that satisfies
\begin{equation}\label{eq:majorana_translation_action}
    T_\chi \chi_{x,y} T_\chi^{-1} = \chi_{x+1,y}.
\end{equation}
(The symmetry generated by Majorana translations in $x$ and in $y$ directions is anomalous in $2+1$d~\cite{HHG160408591}.)

SPTs protected by lattice translations can be constructed by layering one-dimensional lower SPTs in a lattice-translation-symmetric way~\cite{HSH170509243, ET181010539}. 
In this case, there are two Majorana translation SPTs which, respectively, correspond to layering one of the two ${1+1}$d invertible fermionic phases in the $x$-direction. Since the ${1+1}$d Hamiltonians~\eqref{eq:invertible_majorana_chain_hamiltonians} realize these two phases, two SPT Hamiltonians for Majorana translations in the $x$-direction are
\begin{align}
H^\mathrm{maj}_1  
&= i \sum_{x=1}^{L_x} \sum_{\ell=1}^{L_y/2}
\chi_{x,2\ell} \,\chi_{x,2\ell+1} ,\nonumber\\
H^\mathrm{maj}_2  
&= i \sum_{x=1}^{L_x} \sum_{\ell=1}^{L_y/2}
\chi_{x,2\ell-1} \,\chi_{x,2\ell}.
\label{eq:majorana_translation_spt_hamiltonians}
\end{align}
These Hamiltonians are in different ${2+1}$d SPT phases because the ${1+1}$d Hamiltonians~\eqref{eq:invertible_majorana_chain_hamiltonians} are in different ${1+1}$d invertible phases. 
Furthermore, two copies of $H^\mathrm{maj}_1$ is in the same SPT phase as two copies of $H^\mathrm{maj}_2$, which follows from the similar property of the ${1+1}$d Hamiltonians~\eqref{eq:invertible_majorana_chain_hamiltonians}.
The interface of $H^\mathrm{maj}_1$ and $H^\mathrm{maj}_2$ captures the anomaly of ${1+1}$d Majorana translations that manifests through a projective algebra between one-dimensional Majorana translations and fermion number parity~\cite{RZF150503966,HHG160408591,SS230702534}.

\subsection{${2+1}$d SPTs with Majorana translations and CRT}

We consider the same ${2+1}$d lattice system from \appref{2dMajTranlSPT} with Majorana translations~\eqref{eq:majorana_translation_action} but now add a CRT symmetry. In particular, consider the antiunitary $\mathbb Z^\mathrm{CRT}_2$ reflection operator $\Pi$ satisfying
\begin{equation}
    \Pi \chi_{x,y} \Pi^{-1} = (-1)^y\chi_{-x+1,y},\qquad
   \Pi\, i \, \Pi^{-1} = -i. 
\end{equation}
The operators $T_\chi$ and $\Pi$ form the group ${\mathbb Z_{L_x}\rtimes \mathbb Z_2^\mathrm{CRT}}$.

The Hamiltonians~\eqref{eq:majorana_translation_spt_hamiltonians} commute with $\Pi$. Thus, since $H^\mathrm{maj}_1$ and $H^\mathrm{maj}_2$ are in different $\mathbb Z_{L_x}$ Majorana translation SPT phases, they are also in different ${\mathbb Z_{L_x}\rtimes \mathbb Z_2^\mathrm{CRT}}$ SPT phases.
The addition of the $\mathbb Z_2^\mathrm{CRT}$ enhances the ${1+1}$d anomaly order---and hence the ${2+1}$d SPT order---from two to eight~\cite{SSZ250817115}. In particular, two copies of $H^\mathrm{maj}_1$ is in a different ${\mathbb Z_{L_x}\rtimes \mathbb Z_2^\mathrm{CRT}}$ SPT phase from two copies of $H^\mathrm{maj}_2$.

\section{Spectral flow for the lattice axial $\mathbb{R}$ symmetry}
\label{sec:spectral_flow}

Consider the staggered fermion Hamiltonian, which corresponds to \eqref{eq:general_onsager_symmetric_hamiltonian} with ${g_n =\delta_{n,1}}$, and its continuum description as a free massless Dirac fermion field theory.
The two lattice operators $Q_0$, ${\widetilde Q_1 = {Q_1+Q_{-1}\over2}}$ from the Onsager symmetry flow to the vector and axial charges $\mathcal{Q}_\text{V}, \mathcal{Q}_\text{A}$,  respectively, of the Dirac fermion field theory \cite{CPS240912220}. 
Unlike the quantized charge $Q_1$, the charge $\widetilde Q_1$ does not have quantized eigenvalues and generates an  $\mathbb R$ symmetry on the lattice. 
However, it commutes with $Q_0$.

Even though there is no lattice anomaly for the U(1)$\times \mathbb{R}$ symmetry as discussed in \secref{U(1)xR section}, in this appendix we show how the chiral anomaly of the Dirac fermion field theory emerges in the continuum limit for the special case of the staggered fermion Hamiltonian. 
Specifically, since ${[\widetilde Q_1,Q_0]=0}$,  we can study the eigenvalues of $\widetilde Q_1$ in the presence of a $e^{i\theta Q_0}$ symmetry defect and show that in the IR, they exactly match the continuum spectral flow formula for the $\mathcal{Q}_\text{A}$ operator with a $e^{i\theta \mathcal{Q}_\text{V}}$ symmetry defect.
Spectral flow is the hallmark of the 't Hooft anomaly between the vector and axial symmetries in the continuum. 
This lattice spectral flow is only meaningful thanks to the CPT symmetry: the constant term in $\widetilde Q_1$ is fixed by enforcing that the operator anti-commutes with $\Pi$, i.e., 
\begin{equation}
\Pi \widetilde Q_1 \Pi^{-1}  = -\widetilde Q_1\,.
\end{equation}
Without imposing CPT, such a constant shift could be absorbed into a redefinition of $\widetilde Q_1$ in the twisted sector and would not be meaningful. 

\subsection{Spectral flows in CFT}
We begin with a brief review of spectral flows in CFTs with a U(1) global symmetry \cite{SS1987}. 
See, for example, \cite{LS190404833,BOS200302844,CS221112543} for modern discussions in the context of anomalies. 
The spin-1 currents for the U(1) global symmetry have a holomorphic component $J(z)$ and an antiholomorphic component $\overline{J}(\overline{z})$. Their OPEs are
\begin{align}
J(z) J(0) \sim \frac{k}{z^2} ,~~~~
\overline{J}(\overline{z}) \overline{J}(0) \sim \frac{\overline{k}}{\overline{z}^2}.
\end{align}
The 't Hooft anomaly of this $\mathrm U(1)$ symmetry is captured by ${k-\overline{k}}$, which is an integer for fermionic CFTs and an even integer for bosonic CFTs. The zero modes of $J(z)$ and $\overline{J}(\overline{z})$, denoted $J_0$ and $\overline{J}_0$, are the (not necessarily quantized) charges of the left and right-moving components of this $\mathrm U(1)$ symmetry. 
States in the twisted Hilbert space have the following conformal data:
\begin{subequations}\label{eq:cft_twisted_sector_data}
\begin{align}
L_0^\theta &=  L_0  - \frac{\theta}{2\pi} J_0 +  \frac k2\left(\frac{\theta}{2\pi}\right)^2 , \\ 
J_0^\theta &= J_0-  \frac{\theta}{2\pi} k,\\
\overline{L}_0^\theta &=\overline{L}_0  +\frac{\overline{\theta}}{2\pi} \overline{J}_0 +  \frac{\overline{k}}2\left(\frac{\overline{\theta}}{2\pi}\right)^2 , \\
\overline{J}_0^\theta &=\overline{J}_0+  \frac{\overline{\theta}}{2\pi} \overline{k}.
\end{align}
\end{subequations}
Here $\theta $ and $\overline{\theta} $ are the twist angles for $J_0$ and $\overline{J}_0$,  respectively. 
Twisting by the original U(1) global symmetry corresponds to choosing ${\theta=\overline{\theta}}$.

As an example, consider a free massless Dirac fermion field theory and denote its left- and right-moving Weyl fermions by $\Psi_\text{L}$ and $\Psi_\text{R}$, respectively. 
It has a U(1)$_\text{L}\times \text{U(1)}_\text{R}$ global symmetry whose charges are given by
\begin{align}
J_0 = \mathcal{Q}_\text{L} = : \Psi_\text{L}^\dag \Psi_\text{L}:, \quad \overline{J}_0 = \mathcal{Q}_\text{R} = : \Psi_\text{R}^\dag \Psi_\text{R}:
\end{align}
In this case, ${k=\overline{k}=1}$ and $\mathcal{Q}_\text{L}, \mathcal{Q}_\text{R}$ are both quantized. 
The U(1)$_\text{L}$ and U(1)$_\text{R}$ symmetries are both  anomalous. However, each of the vector and axial symmetries U(1)$_\text{V}$ and U(1)$_\text{A}$ generated by
\begin{align}
\mathcal{Q}_\text{V} = J_0 + \overline{J}_0 ,~~~\mathcal{Q}_\text{A}  = J_0-\overline{J}_0,
\end{align}
is individually  anomaly-free.\footnote{For the staggered fermion Hamiltonian \eqref{eq:antiperiodic_staggered_hamiltonian}, while there is a lattice U(1) symmetry generated by $Q_0$ ($Q_1$) flowing to U(1)$_\text{V}$ (U(1)$_\text{A}$), there cannot be a lattice U(1) symmetry for U(1)$_\text{L}$ or U(1)$_\text{R}$  because of the anomaly \cite{KS240102533}. Even though there is a lattice operator $Q_0+Q_1\over2$ flowing to $\mathcal{Q}_\text{L}$, it is not quantized and generates an $\mathbb{R}$ symmetry. Alternatively, there is an $S^1$ family of  conserved lattice operators $\exp(i \theta Q_0/2 )\exp(i \theta Q_1/2)$ (i.e., $\theta$ is $2\pi$-periodic) which flows to $\exp( i \theta \mathcal{Q}_\text{L})$, but they do not form a U(1) group under multiplication on the lattice. (Here we have used $e^{i \pi Q_0 } = e^{i \pi Q_1}$ being the order-2 fermion parity.)}

The spectral flow formula for the scaling dimension ${\Delta=  L_0+\overline{L}_0}$ and the axial charge $\mathcal{Q}_\text{A}$ in the $\mathcal{Q}_\text{V}$-twisted Hilbert space (${\overline{\theta}  = \theta }$) are
\begin{subequations}\label{eq:continuum_spectral_flow_data}
\begin{align}\label{eq:vector_twisted_spectral_flow} 
\Delta_\theta &= \Delta  +\left(\frac{\theta}{2\pi}\right)^2- \frac{\theta}{2\pi} \mathcal{Q}_\text{A}   \\
\mathcal{Q}_{\text{A},\theta} &=  \mathcal{Q}_\text{A} - \frac{\theta}\pi \,.
\end{align}
On the other hand, the $\mathcal{Q}_\text{A}$-twisted Hilbert space (${\overline{\theta}  = - \theta }$) has the following conformal data   
\begin{align}
\Delta_\theta  &= \Delta  + \left(\frac{\theta }{2 \pi}\right)^2  - \frac{\theta}{2 \pi }\mathcal{Q}_\text{V}, \\  \mathcal{Q}_{\text{V}, \theta } & = \mathcal{Q}_\text{V} - \frac{\theta }{ \pi } \,.
\end{align}
\end{subequations}

\subsection{Spectral flows on the lattice}

We now turn to the staggered fermion Hamiltonian. 
To avoid complications from the zero modes, in this appendix alone, we work with antiperiodic boundary conditions. This simply means we take the opposite sign for the coupling between $c_L$ and $c_{L+1}$ so that the Hamiltonian and $\widetilde Q_1$ charge are 
\begin{subequations}
\begin{align}
H 
&=  - i \sum_{j=1}^{L-1} c_j^\dagger c_{j+1} 
 + i  c_L^\dagger c_1 +\text{h.c.}, \label{eq:antiperiodic_staggered_hamiltonian} \\
\widetilde Q_{1} 
&=\frac12 \sum_{j=1}^{L-1} c_j^\dagger c_{j+1}
 - \frac 12 c_L^\dagger c_1 +\text{h.c.}
\end{align}
\end{subequations}
As usual with anti-periodic boundary conditions, it is convenient to define momenta to be half-integer
\begin{align}
&c_j  =\frac{1}{\sqrt{L}} 
\sum_{k =  \frac{1}{2}}^{L- \frac{1}{2}} e^{\frac{ 2 \pi i }{L}jk} \gamma_k, \quad 
c_j^\dagger  =\frac{1}{\sqrt{L}} 
\sum_{k =  \frac{1}{2}}^{L  - \frac{1}{2}} e^{-\frac{2 \pi i }{L}jk} \gamma_k^\dagger.
\end{align}
Consider inserting a $\mathrm U(1)_\text{V}$ defect with angle $\theta$ at link $(0,1)$. The modified Hamiltonian and charge are  
\begin{subequations}
\begin{align}
H_\theta 
&=  - i \sum_{j=1}^{L-1} c_j^\dagger c_{j+1} 
+ i e^{- i \theta} c_L^\dagger c_1 +\text{h.c.}, \\
\widetilde Q_{1 ,\theta} 
&=\frac12 \sum_{j=1}^{L-1} c_j^\dagger c_{j+1}
 -\frac 12 e^{-i\theta}c_L^\dagger c_1 +\text{h.c.}
\end{align}
\end{subequations}
We can perform a unitary transformation $c_j \mapsto  e^{- \frac{ ij \theta }{L}}c_j$ to spread out the defect across the entire chain. 
In this unitary frame, the Hamiltonian and the unquantized charge are 
\begin{subequations}
\begin{align}
H_\theta 
&=  - i \sum_{j =1}^{L-1} e^{-\frac{i \theta}{L}}c_j^\dagger c_{j+1}
+ i e^{-\frac{i \theta}{L}}c_L^\dagger c_{1} 
 +\text{h.c.}\nonumber\\
&=  2 \sum_{\mathclap{k=\frac 12}}^{ L - \frac{1}{2}} 
\sin \left(\frac{2 \pi k - \theta }{L}\right)
\gamma_k^\dagger \gamma_k, \\
\widetilde Q_{1 ,\theta} 
&= \frac12 \sum_{j =1}^{L-1} 
 e^{-\frac{i \theta}{L}}c_j^\dagger c_{j+1} - \frac{1}{2}e^{-\frac{i \theta}{L}}c_L^\dagger c_{1} 
+\text{h.c.}\nonumber\\
&=  \sum_{\mathclap{k =  \frac{1}{2}}}^{L - \frac{1}{2}}
\cos \left(\frac{2 \pi k- \theta }{L}\right)
\gamma_k^\dagger \gamma_k .\label{eq:tilde_q_one_momentum_space} 
\end{align}
\end{subequations}

Let us normal order these operators and examine their low-energy excitations. 
The ground state $\left|  \Omega \right\rangle$ when $\theta=0$ with antiperiodic boundary conditions is unique, and is defined by
\begin{equation}\begin{aligned}
&\gamma_k \left|  \Omega \right\rangle =  0,~~~~0< k <\frac L2 , \\
&\gamma_k^\dagger \left|  \Omega \right\rangle =  0,~~~~-\frac L2< k <0 .
\end{aligned}\end{equation}
To normal order $H_\theta$, we write it as
\begin{align}
H_\theta  
= 2 \sum_{k=\frac 12}^{\frac L2- \frac{1}{2}} 
&\sin \left(\frac{2 \pi k - \theta }{L}\right)
\gamma_k^\dagger \gamma_k\nonumber\\
+2 \sum_{ \mathclap{k=\frac{L}{2} + \frac 12}}^{L- \frac{1}{2}} 
&\sin \left(\frac{2 \pi k - \theta }{L}\right)
\left(-\gamma_k^\dag \gamma_k +1\right) 
\end{align}
The normal ordering constant can be evaluated with standard trigonometric identities. The result is 
\begin{align}
H_\theta    
= 2 \sum_{k=\frac 12}^{\frac L2- \frac{1}{2}} 
\Biggl[
&\sin \left(\frac{2 \pi k - \theta }{L}\right)
\gamma_k^\dagger \gamma_k\nonumber\\
+& \sin \left(\frac{2 \pi k + \theta }{L}\right) 
 \gamma_{-k} \gamma_{-k}^\dag \Biggr]\\
-2&
\cos \left(\frac\theta L\right) \csc \left( \frac{\pi}{L}\right)\nonumber
\end{align}

To compare with the CFT data, we rescale the Hamiltonian by  ${E_\theta = \frac{L}{4\pi} H_\theta}$.  
The low-lying states and their large $L$ limits are listed in \eqref{eq:lattice_spectral_flow_energies}, where we discard the non-universal term linear in $L^2$ (due to the possibility of adding a constant potential on each site in $H_\theta $).
The first four excited states at low energy correspond to the operators ${\Psi_\text{L}, \Psi_\text{R} , \Psi_\text{L}^\dagger, \Psi_\text{R}^\dagger}$ of the CFT via the operator-state correspondence.
These have scaling dimension ${\Delta =\frac{1}{2}}$ and their $\mathcal{Q}_\text{A}$ eigenvalues are $-1,+1,+1,-1$, respectively. 
Up to the unphysical term that is linear in $L^2$ in $E$, the energies are related to the scaling dimension by ${E_\theta= \Delta_\theta - \frac c{12}}$. These match perfectly with the spectral flow formula in the CFT \eqref{eq:vector_twisted_spectral_flow} upon identifying $\mathcal{Q}_\text{A}$ as the continuum limit of $\widetilde Q_1$. 

\addtocounter{equation}{1}

Next, consider the unquantized axial charge in the presence of the $\mathrm U(1)_\text{V}$ defect. Normal ordering \eqref{eq:tilde_q_one_momentum_space} we find 
\begin{align}
\widetilde Q_{1, \theta} 
= \sum_{k = \frac{1}{2}}^{\frac{L}{2} - \frac{1}{2}}
\Biggl[
&\cos \left(\frac{2 \pi k - \theta }{L}\right)
\gamma_k^\dagger \gamma_k\nonumber\\
-&\cos\left( 
\frac{2\pi k +\theta}{L}
\right)\gamma_{-k} \gamma_{-k}^\dag\Biggr]\\
-& \sin \left(\frac{\theta}{L} \right)
\csc \left(\frac{ \pi}{L}\right).\nonumber
\end{align}
In the large $L$ limit, the effect of the $\mathrm U(1)_\text{V}$ defect is to shift all the $\widetilde Q_1$ eigenvalues uniformly by 
\begin{align}
\widetilde Q_{1,\theta}  = \widetilde Q_1 - \frac{\theta}\pi +\cdots . 
\end{align}
Again, this matches the formula for spectral flow from the CFT \eqref{eq:vector_twisted_spectral_flow}. 
\begin{widetextnorules}
\newcommand{\Efourteenrowsep}{0.8ex}
\setlength{\arraycolsep}{5pt}
\renewcommand{\arraystretch}{2.4}
\begin{equation}
\begin{array}{@{}r@{\qquad}r@{\quad\approx\quad}l@{}}
\text{State} & E_\theta & E_\theta \big|_{L \gg 1} \\
\hline
\left|  \Omega \right\rangle
&
\displaystyle
-\frac{L}{2\pi}  \cos\left(\frac\theta L\right)
\csc\left( \frac{\pi}{L}\right)
&
\displaystyle
-\frac{1}{12} + 
\left( \frac{\theta}{2\pi}\right)^2 + \cdots 
\\[\Efourteenrowsep]
\gamma_{-\frac12}\left|  \Omega \right\rangle
&
\displaystyle
\frac{L}{2\pi} \sin \left(\frac{\pi+\theta}{L}\right)
-\frac{L}{2\pi}  \cos\left(\frac\theta L\right)
\csc\left( \frac{\pi}{L}\right)
&
\displaystyle
-\frac{1}{12} + \left( \frac{\theta}{2\pi}\right)^2 + 
\frac12 +\frac{\theta}{2\pi}
+\cdots
\\[\Efourteenrowsep]
\gamma_{-(\frac L2-\frac12)}\left|  \Omega \right\rangle
&
\displaystyle
\frac{L}{2\pi} \sin \left(\frac{\pi-\theta}{L}\right)
-\frac{L}{2\pi}  \cos\left(\frac\theta L\right)
\csc\left( \frac{\pi}{L}\right) 
&
\displaystyle
-\frac{1}{12} 
 + \left( \frac{\theta}{2\pi}\right)^2
+ \frac12 -\frac{\theta}{2\pi}+\cdots
\\[\Efourteenrowsep]
\gamma^\dagger_{\frac12}\left|  \Omega \right\rangle
&
\displaystyle
\frac{L}{2\pi} \sin \left(\frac{\pi-\theta}{L}\right)
-\frac{L}{2\pi}  \cos\left(\frac\theta L\right)
\csc\left( \frac{\pi}{L}\right)
&
\displaystyle
-\frac{1}{12} 
+  \left( \frac{\theta}{2\pi}\right)^2
+ \frac12 -\frac{\theta}{2\pi}+\cdots
\\[\Efourteenrowsep]
\gamma^\dagger_{\frac L2-\frac12}\left|  \Omega \right\rangle
&
\displaystyle
\frac{L}{2\pi} \sin \left(\frac{\pi+\theta}{L}\right)
-\frac{L}{2\pi}  \cos\left(\frac\theta L\right)
\csc\left( \frac{\pi}{L}\right)
&
\displaystyle
-\frac{1}{12} 
+ \left( \frac{\theta}{2\pi}\right)^2
+ \frac12 +\frac{\theta}{2\pi}+\cdots
\end{array}
\tag{E14}\label{eq:lattice_spectral_flow_energies}
\end{equation}
\end{widetextnorules}

\bibliographystyle{ytphys}
\bibliography{refs}

\end{document}